\documentclass{optica-article}

\journal{opticajournal} 

\articletype{Research Article}
\usepackage{braket}
\usepackage{lineno}
\usepackage{siunitx}

\usepackage{fancyhdr}
\begin{document}

\title{Coherent Feedback Cooling of an Ultracoherent Phononic-Crystal Membrane at Room Temperature}

\author{Luiz Couto Correa Pinto Filho,\authormark{1,2,*,\textdagger}
        Yingxuan Chen,\authormark{1,*}
        Frederik Werner Isaksen,\authormark{1}
        Daniel Allepuz-Requena,\authormark{1}
        Angelo Manetta,\authormark{1}
        Dennis Henneberg Høj,\authormark{1}
        Ulrich Busk Hoff,\authormark{1}
        Alexander Huck,\authormark{1}
        and Ulrik Lund Andersen\authormark{1,\textdaggerdbl}}

\address{\authormark{1}Center for Macroscopic Quantum States (bigQ), Department of Physics, Technical University of Denmark, Fysikvej, Kongens Lyngby, Denmark\\
\authormark{2}Danish Fundamental Metrology (DFM), Kogle Alle 5, DK-2970 Hørsholm, Denmark}

\address{\authormark{*}These authors contributed equally to this work.}
\email{\authormark{\textdagger}lco@dfm.dk}
\email{\authormark{\textdaggerdbl}andersen@fysik.dtu.dk}

\begin{abstract*} 
Optomechanical systems provide a versatile platform for precision measurements and investigations of fundamental physics, where bringing macroscopic resonators into the quantum regime is a widely pursued goal. Achieving such quantum behavior of solid-state mechanical resonators at room temperature would greatly broaden their applications by removing the need for cryogenic environments. Reaching this goal requires efficient cooling of mechanical motion, among various laser cooling methods, dynamical backaction cooling (DBC) is widely utilized in experiments but fundamentally limited when operating in the sideband-unresolved regime. Coherent feedback cooling (CFC) can overcome this limitation, while avoiding state collapse and the electronic restrictions inherent to measurement-based feedback. Here, we experimentally demonstrate CFC using an ultracoherent density phononic crystal membrane. By combining CFC with strong DBC in a relatively narrow cavity, we achieve a phonon occupation reduction from $5.5\times10^{6}$ to $166\pm7$, corresponding to a cooling factor of $3.3\times10^{4}$ at room temperature, even with current experimental limitations. Our results show the potential of CFC for approaching the ground state of high-$Q$ membranes at room temperature.
\end{abstract*}

\section{Introduction}
Optomechanical systems\cite{bowen2015quantum,aspelmeyer2014cavity,barzanjeh2022optomechanics}, which describe the interaction between electromagnetic fields and mechanical resonators, have attracted significant interest in quantum optics and precision measurement. They serve as versatile platforms for fundamental quantum physics, enabling ground-state cooling of macroscopic resonators\cite{o2010quantum,teufelSidebandCoolingMicromechanical2011,chanLaserCoolingNanomechanical2011,wilson2015measurement,clark2017sideband,rossiMeasurementbasedQuantumControl2018,delic2020cooling,tebbenjohanns2021quantum,magrini2021real,dania2025high}, generation of squeezed light\cite{brooks2012non,safavi2013squeezed,purdy2013strong,nielsen2017multimode,mason2019continuous,aggarwal2020room,magrini2022squeezed,schmidSqueezingLightOptomechanical2025}, and mechanical squeezing\cite{lecocq2015quantum,pirkkalainen2015squeezing,wollman2015quantum,szorkovszky2011mechanical}. Optomechanical systems are also widely used for sensing, including force sensing\cite{gavartinHybridOnchipOptomechanical2012,xia2023entanglement}, gravitational wave detection\cite{abramovici1992ligo,abbott2016observation,whittle2021approaching}, magnetic field sensing\cite{forstner2012cavity,yu2016optomechanical}, and accelerometry\cite{krause2012high}. All of these applications strongly rely on suppressing the thermal phonon occupation of the mechanical resonator. Ground-state cooling has already been demonstrated in several different platforms\cite{o2010quantum,teufelSidebandCoolingMicromechanical2011,chanLaserCoolingNanomechanical2011,wilson2015measurement,clark2017sideband,rossiMeasurementbasedQuantumControl2018,delic2020cooling,tebbenjohanns2021quantum,magrini2021real,dania2025high}. However, for solid-state mechanical resonators, cryogenic precooling is usually required to observe their quantum behavior\cite{o2010quantum,teufelSidebandCoolingMicromechanical2011,chanLaserCoolingNanomechanical2011,safavi2013squeezed,purdy2013strong,wilson2015measurement,lecocq2015quantum,pirkkalainen2015squeezing,wollman2015quantum,clark2017sideband,nielsen2017multimode,moller2017quantum,rossiMeasurementbasedQuantumControl2018,riedinger2018remote,mason2019continuous,galinskiy2020phonon,chen2020entanglement,thomas2021entanglement}. Reducing or eliminating the reliance on cryogenic temperatures would significantly simplify the experimental setup and broaden the range of applications, motivating the development of systems and methods capable of bringing macroscopic resonators into the quantum regime at room temperature\cite{saarinen2023laser,huangRoomtemperatureQuantumOptomechanics2024a,xia2025motional}.

Ultracoherent membrane resonators engineered with density phononic crystal patterning have achieved record quality factors of $\sim 1\times10^{9}$ and $Qf$ products of $\sim 1\times10^{15}$ at room temperature\cite{hoj2024ultracoherent}. These membranes have already enabled pondermotive squeezing\cite{huangRoomtemperatureQuantumOptomechanics2024a,schmidSqueezingLightOptomechanical2025}, underlining their strong potential for ground-state cooling at room temperature. 

Among the available laser-cooling schemes, cavity dynamical backaction cooling (DBC) is the most widely used in experiments\cite{purdy2013strong,nielsen2017multimode,mason2019continuous,galinskiy2020phonon,saarinen2023laser,huangRoomtemperatureQuantumOptomechanics2024a,xia2025motional}. However, its cooling effect is fundamentally limited in the sideband-unresolved regime, where the cavity linewidth is larger than the mechanical frequency\cite{aspelmeyer2014cavity}. In this regime, measurement-based feedback cooling can provide more effective suppression of thermal fluctuations by measuring the optical output spectrum and feeding back a classical signal with carefully chosen frequency range, bandwidth, and phase\cite{wiseman2009quantum,wilson2015measurement,rossiMeasurementbasedQuantumControl2018,whittle2021approaching,tebbenjohanns2021quantum}. Nevertheless, the measurement process inevitably collapses the quantum state\cite{rossiObservingVerifyingQuantum2019} and adds backaction noise\cite{rossiMeasurementbasedQuantumControl2018,wilson2015measurement}. 

 Coherent feedback cooling (CFC) \cite{hamerly2012advantages,hamerly2013coherent,yamamoto2014coherent,jacobs2014coherent,wang2017enhancing,li2017enhanced,harwood2021cavity,guo2022coherent,schmid2022coherent,bemani2024optical} is achieved by re-injecting the optical probe field into the system after an engineered delay and an optional displacement operation, without intermediate measurement. This scheme preserves the quantum coherence of the system, avoids measurement-induced backaction noise and the technical restrictions induced by electronic devices in measurement-based feedback loops. Coherent feedback control has been applied to tasks such as noise cancellation\cite{tsang2010coherent}, entanglement generation\cite{zhou2015quantum}, and squeezing enhancement\cite{gough2009enhancement}. Recently, Ernzer \textit{et al.} \cite{ernzer2023optical} demonstrated the first all-optical CFC of a nanomechanical membrane operating in the deep sideband-unresolved regime at cryogenic temperatures.

Differently from \cite{ernzer2023optical}, here we report the first demonstration of CFC combined with DBC in a near-sideband-resolved cavity at room temperature, enabled by a high-$Q$ macroscopic membrane resonator with a density phononic crystal design. We develop a model beyond the fast-cavity approximation that accounts for finite cavity bandwidth and incorporates experimentally relevant frequency fluctuations. With this system, we reduce the phonon occupation of the mechanical mode at $\Omega_{\rm m}=2\pi\times1.14$ \si{\mega \hertz} from $5.5\times10^{6}$ to $166 \pm 7$, corresponding to an effective temperature of $\SI{9.1}{\milli\kelvin}$. Currently limited by relatively weak optomechanical coupling and feedback-loop loss, we nonetheless achieve a cooling factor of $3.3\times10^4$. These results demonstrate CFC 
as a highly effective technique for cooling ultracoherent mechanical resonators and advance the long-term goal of ground-state operation without cryogenics, while highlighting the synergistic action of dynamical backaction and coherent feedback.

\section{Conceptual Principle}

\begin{figure}
    \centering
    \includegraphics[width=1\linewidth]{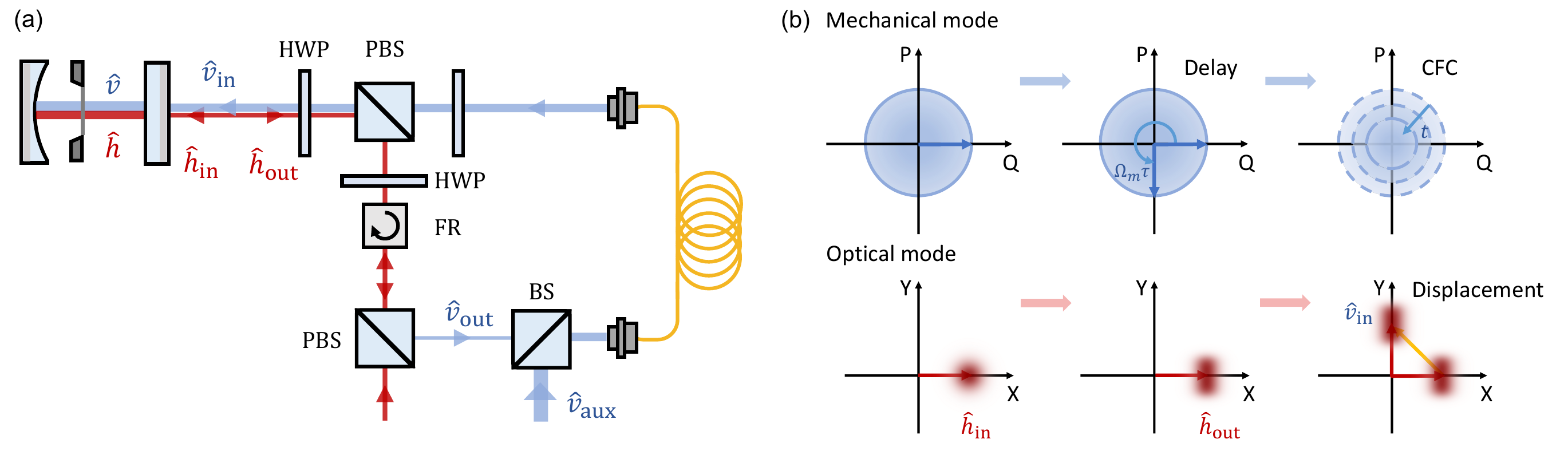}
    \captionsetup{width=1\linewidth}
    \caption{Conceptual scheme of CFC. (a)	Experimental schematic. The probe beam $\hat{h}_{\rm{in}}$ is sent into the optomechanical cavity. After the optomechanical interaction between the horizontally polarized intracavity field $\hat{h}$ and the membrane, the reflected output beam $\hat{h}_{\rm{out}}$ is converted to vertical polarization and combined with a strong auxiliary beam $\hat{v}_{\rm{aux}}$ on a beam splitter (BS) where a displacement operation is implemented to convert optical phase information into amplitude. The resulting field is then sent through a fiber delay and re-injected into the cavity, providing a coherent feedback force on the membrane. $\hat{v}_{\rm{in}}$ and $\hat{v}$ denote the annihilation operators of the input cooling beam and vertically polarized intracavity field, respectively. HWP, half-wave plate; PBS, polarizing beam splitter; FR, Faraday rotator. (b)	Phase-space representation of the CFC process. The mechanical mode (blue) evolves from a thermal state with the uncertainty circle in the initial phase-space plot to a reduced uncertainty area over time under CFC. $\Omega_{\rm{m}}\tau$ is the duration associated with the fiber delay, during which the $\hat Q$ quadrature information evolves into its $\hat P$ quadrature. The optical mode (red) sequentially experiences phase modulation by the mechanical motion, a displacement operation converting phase to amplitude modulation, and reinjection into the cavity with an amplitude modulation proportional to negative momentum of the resonator. $\hat Q$ and $\hat P$ denote the position and momentum quadratures of the mechanical resonator, respectively. $\hat X$ and $\hat Y$ denote the amplitude and phase quadratures of the optical field, respectively. The phase-space representation is plotted for the on-resonance condition.
}
    \label{concept}
\end{figure}

We introduce the schematic for the CFC protocol in Fig. \ref{concept}(a), and the corresponding phase-space evolution of the optical and mechanical modes in Fig. \ref{concept}(b). For clarity, we first consider an idealized lossless feedback loop and assume a resonant cavity drive so that the output field does not undergo additional quadrature rotation. We consider a mechanical mode $\hat{b}$ with position and momentum quadratures $\hat{Q}$ and $\hat{P}$, obeying $[\hat{Q},\hat{P}]=i$. At room temperature the resonator is in a large thermal state, sketched as a blue uncertainty circle in Fig.~\ref{concept}(b). A coherent optical probe field $\hat{h}_{\rm in}$, represented as a displaced vacuum state with quadratures $\hat{X}$ and $\hat{Y}$, impinges on the optomechanical cavity. Through the optomechanical interaction, the displacement ($\hat{Q}$) of the membrane is imprinted on the phase quadrature ($\hat{Y}_{\rm{out}}$) of the reflected beam $\hat{h}_{\rm{out}}$, which is represented as a modulation in the optical phase of the sideband at the mechanical frequency. From a time-averaged perspective, the modulation can be regarded as a small oscillation of the optical mode along the phase quadrature ($\hat{Y}_{\rm{out}}$), which smears the Gaussian uncertainty distribution into an elliptical shape.

Instead of measuring the phase quadrature with homodyne detection and feeding classical information back based on measurement results, as done in conventional feedback cooling experiments, here we send the optical field back in an orthogonal polarization mode $\hat{v}_{\rm{in}}$ to actuate the resonator. The optomechanical interaction couples the optical amplitude quadrature ($\hat{X}$) to the mechanical momentum ($\hat{P}$) allowing the radiation pressure force to drive the mechanical displacement ($\hat{Q}$). However, after the first pass in the cavity, the mechanical displacement information is imprinted on the optical phase quadrature ($\hat{Y}$). Therefore, a displacement operation (see Supplement~1 for more details) is required to convert the modulation in the phase of the optical field into amplitude ($\hat{X}$). This is implemented by rotating the polarization from $\hat{h}_{\rm{out}}$ to $\hat{v}_{\rm{out}}$ and combining it with a strong auxiliary field $\hat{v}_{\rm{aux}}$ in an asymmetric beam splitter (BS). An optimal displacement angle $\gamma$ can be chosen to achieve the most efficient conversion from phase fluctuation to amplitude fluctuation. To make the feedback scheme work effectively for cooling, a delay $\Omega_{\rm{m}}\tau$ corresponding to a phase space rotation of the mechanical quadrature at $\Omega_{\rm{m}}$ from $\hat{Q}$ to $-\hat{P}$ is introduced, allowing the mechanical displacement information to evolve into momentum information. In this way, the reinjected optical field $\hat{v}_{\rm{in}}$ provides a feedback force that effectively increases the damping rate leading to coherent cooling. As a result, the uncertainty area of the mechanical mode in phase space decreases continuously over time.

\section{Theoretical Model}

The theoretical model of the CFC scheme is described in this section. In the optomechanical cavity, we consider that the mechanical mode $\hat{b}$, with mechanical resonance frequency $\Omega_{\rm{m}}$, is coupled to the two optical intracavity modes $\hat{h}$ and $\hat{v}$ of horizontal and vertical polarization, respectively. 
We assume that the bosonic mode operators $\hat{a}$ obey the canonical commutation relation 
$[\hat{a},\hat{a}^{\dagger}] = 1$, while the input field operators $\hat{a}_{\mathrm{in}} = \hat{a}_{\mathrm{in}}(t)$ obey the standard commutation relation $[\hat{a}_{\mathrm{in}}(t),\hat{a}^{\dagger}_\mathrm{in}(t')] = \delta(t-t')$.

The input field $\hat{h}_{\mathrm{in}}$ is coupled into a cavity with linewidth $\kappa$ through a mirror with coupling rate $\kappa^{\mathrm{in}}$. The corresponding input-output relation is given by $\hat{h}_{\rm{out}}(t)=-\hat{h}_{\rm{in}}(t)+\sqrt{\kappa^{\rm{in}}}\hat{h}(t)$, where $\hat{h}_{\rm{in}}$, $\hat{h}$, and $\hat{h}_{\rm{out}}$ are the annihilation operators for the input field, intra-cavity field, and output field of the horizontal polarization mode, respectively. The output mode then observes a phase shift of angle $\phi$, a displacement of complex amplitude $\delta_{0}$, and a delay  of  duration $\tau$ before it is sent back into the cavity as a vertical polarization mode. The resulting vertical polarization input field is given by 
\begin{equation}
\hat{v}_{\mathrm{in}} = \sqrt{\eta}T_{\tau}e^{-i\phi}\hat{h}_{\mathrm{out}} + \delta + \sqrt{1-\eta}\hat{h}_{v},
\end{equation}
where $\eta$ characterizes the efficiency of the feedback loop, considering the optical losses experienced after leaving the cavity. $T_{\tau}$ denotes a delay-line operator such that for an arbitrary operator $\hat{a}(t)$, $T_{\tau}\hat{a}(t) := \hat{a}(t-\tau)$. The complex displacement is redefined as $\delta = \sqrt{\eta_f}\delta_0$, where $\eta_f$ accounts for optical losses introduced after the displacement operation. $\hat{h}_{v}$ is the horizontal polarization vacuum mode induced by optical losses in the feedback loop. For suitable choices of $\phi$, $\delta$, and $\tau$, the re-injected field carries an amplitude modulation proportional to the mechanical momentum, thereby producing a radiation-pressure force that enhances the mechanical damping.

The system Hamiltonian is expressed as
\begin{equation}
\hat{H} = \hbar\Omega_{\rm m}\hat{b}^{\dagger}\hat{b} + \hbar\Delta_{h}\hat{h}^{\dagger}\hat{h} + \hbar g_{0,h}\hat{h}^{\dagger}\hat{h}(\hat{b}^{\dagger} + \hat{b}) + \hbar\Delta_{v}\hat{v}^{\dagger}\hat{v} + \hbar g_{0,v}\hat{v}^{\dagger}\hat{v}(\hat{b}^{\dagger} + \hat{b}),
\end{equation}
where $\Delta_{h}$ and $\Delta_{v}$ are the detuning of the horizontal and the vertical polarization modes, $g_{0,h}$ and $g_{0,v}$ represent the vacuum optomechanical coupling rates.
In a linearized and rotating frame, the Langevin equations of the system are given by (for clarity, the fluctuation notation is omitted and we define $\delta\hat{O}$ as $\hat{O}$.)
\begin{align}
\dot{\hat{Q}} &= \Omega_{\rm{m}} \hat{P}, \\
\dot{\hat{P}} &= -\Omega_{\rm{m}} \hat{Q} - \Gamma_{\rm{m}} \hat{P} - \sqrt{2}|g_{h}|( e^{-iu}\hat{h} + e^{iu} \hat{h}^{\dagger}) - \sqrt{2}|g_{v}|(e^{-ix}\hat{v} + e^{ix}\hat{v}^{\dagger}) + \sqrt{2\Gamma_{\rm{m}}}\hat{P}_{\text{in}},\\
\dot{\hat{h}} &= - \frac{\kappa}{2}\hat{h} + i\Delta_{h}^{\rm{eff}}\hat{h} + i|\alpha_{h,\rm{cav}}|\delta\Delta_{h} -i\sqrt{2}g_{h}\hat{Q} + \sqrt{\kappa^{\text{in}}}\hat{h}_{\text{in}} + \sqrt{\kappa-\kappa^{\mathrm{in}}}\hat{h}_{\text{loss}},\\
\dot{\hat{v}} &= - \frac{\kappa}{2}\hat{v} + i\Delta_{v}^{\rm{eff}}\hat{v} + i|\alpha_{v,\rm{cav}}|\delta\Delta_{v} -i \sqrt{2}g_v\hat{Q} + \sqrt{\kappa^{\text{in}} \eta}e^{-i\phi}T_{\tau}(\hat{h}_{\text{in}} - \sqrt{\kappa^{\text{in}}}\hat{h}) +\sqrt{\kappa- \eta\kappa^{\mathrm{in}}}\hat{v}_{\mathrm{v}}.
\end{align}
$\hat{Q} = (\hat{b}+\hat{b}^{\dagger})/\sqrt{2}$, $\hat{P} = (i\hat{b}^{\dagger}-i\hat{b})/\sqrt{2}$ are the dimensionless position and momentum quadrature operators of the mechanical mode, respectively. The mechanical damping rate is denoted by $\Gamma_{\rm{m}}$. The effective optomechanical coupling strengths of the horizontal and vertical polarization modes are denoted by $g_h$ and $g_v$, respectively. Both $g_h$ and $g_v$ are complex numbers with phases $\text{Arg}(g_{h}) = u$ and $\text{Arg}(g_{v}) = x$, respectively. $\hat{P}_{\mathrm{in}} = \hat{P}_{\mathrm{in}}(t)$ describes the coupling of the mechanical mode to its thermal environment. With Markovian and non-rotating wave approximation, $\braket{\hat{P}_{\mathrm{in}}(t)} = 0$ and $\braket{\hat{P}_{\mathrm{in}}(t),\hat{P}_{\mathrm{in}}(t')} = (\bar{n}_{\rm in} + \frac{1}{2})\delta(t-t')$, where $\bar{n}_{\rm in}$ denotes the mean thermal occupation number of the bath at temperature $T$. $\hat{h}_{\mathrm{loss}}$ and $\hat{v}_{\mathrm{v}}$ are vacuum modes induced by optical losses. $\Delta_{h}^{\rm{eff}}$ and $\Delta_{v}^{\rm{eff}}$ are mean effective detuning levels, defined as $\Delta_{h}^{\text{eff}} := \Delta_{h}+\braket{\delta\Delta_{h}} - \sqrt{2}g_{0,h}\braket{Q}$ and $\Delta_{v}^{\text{eff}} := \Delta_{v}+\braket{\delta\Delta_{v}} - \sqrt{2}g_{0,v}\braket{Q}$, respectively. The red-detuned configuration is characterized by a negative effective detuning, i.e., $\Delta_{h}^{\rm{eff}}<0$  and $\Delta_{v}^{\rm{eff}}<0$. $\delta\Delta_{h}$ and $\delta\Delta_{v}$ denote fluctuations in the detuning \cite{Huang2024EPFL}, introduced to account for additional frequency fluctuations originating from sources such as mirror vibrations, laser-lock instability, etc. $|\alpha_{h,\rm{cav}}|$ and $|\alpha_{v,\rm{cav}}|$ represent the mean intra-cavity photon numbers of the horizontal polarization mode and vertical polarization mode, respectively.

The detailed derivation of the Quantum Langevin Equations is provided in Supplement~1. By solving these equations, we obtain the analytical expression for the power spectral density (PSD) of the mechanical position, which is defined as
$S_{\hat{Q}\hat{Q}}(\omega) = \frac{1}{2\pi}\int_{-\infty}^{\infty}\braket{\hat{Q}(\omega)\hat{Q}(\omega')}d\omega'$, where $\hat{Q}(\omega) = \int_{-\infty}^{\infty}\hat{Q}(t)e^{i\omega t}dt$ is the Fourier transform of time domain position operator $\hat{Q}(t)$. The phonon occupation $\bar{n} = \braket{\hat{b}^{\dagger}\hat{b}}$ can be obtained by
\begin{equation}\label{eq:phonon_SQQ}
\bar{n} = \frac{1}{4\pi}\int_{-\infty}^{\infty}\left(1+\frac{\omega^2}{\Omega_{\rm m}^2}\right)S_{\hat{Q}\hat{Q}}(\omega)d\omega - \frac{1}{2}.
\end{equation}
An analytical expression for the detected PSD $S_{\hat{Y}_{\rm{det}}\hat{Y}_{\rm{det}}}$ is likewise derived in Supplement~1 and is used to fit the experimental spectra and extract system parameters.

\section{Experimental Setup}

\begin{figure}
    \centering
    \includegraphics[width=1\linewidth]{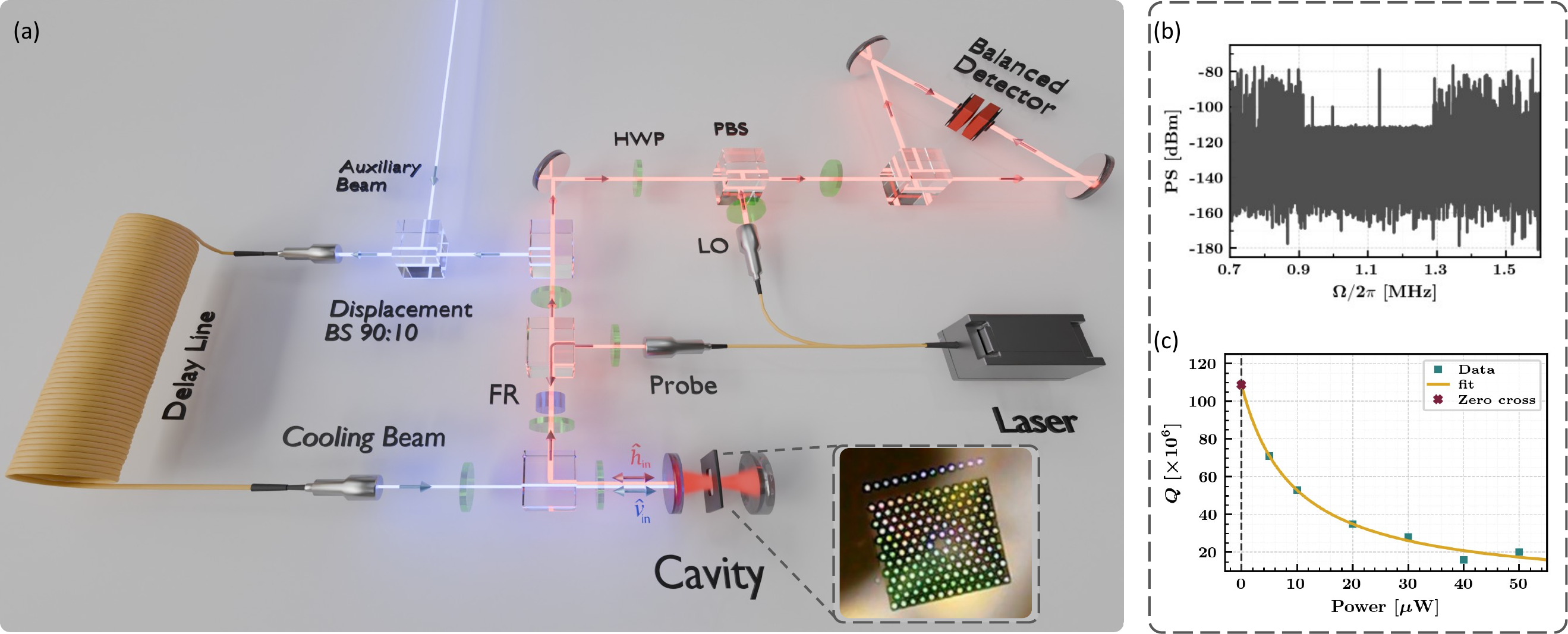}
    \captionsetup{width=1\linewidth}
    \caption{The experiment.
(a) Experimental setup for CFC scheme. A horizontally polarized probe beam is injected into the cavity containing a membrane. The reflected beam is rotated to vertical polarization by using a FR and a HWP. A small portion is directed to a balanced homodyne detector for readout, while the main beam is combined with a strong auxiliary beam on a 90:10 BS for the displacement operation. The resulting field passes through a fiber delay line before being re-injected into the cavity to provide coherent feedback. LO, local oscillator. \textit{Inset}: A camera image of the $\rm{Si_{3}N_{4}}$ membrane showing the phononic-crystal pattern. (b) Phononic-crystal bandgap of the membrane around the fundamental mode at $\Omega_{\rm{m}} = 2\pi\times1.14$ \si{\mega \hertz}, obtained from the power spectrum (PS) of the phase quadrature of the reflected probe beam.
(c) Mechanical ringdown measurement used to characterize the mechanical quality factor $Q$. The measurement is performed at different probe powers under large red detuning to minimize optomechanical damping, and the intrinsic $Q$ is extracted by fitting the measured data and extrapolating to zero probe power.}
    \label{fig:setup}
\end{figure}

The experimental setup for the CFC scheme is shown in Fig. \ref{fig:setup}(a). The optomechanical system consists of a Fabry-Pérot (FP) cavity containing a silicon nitride membrane with density phononic crystal structure in a Membrane-in-the-middle (MIM) configuration. The vibration of the central defect gives rise to the fundamental mechanical resonator mode at $\Omega_{\rm{m}} = 2\pi\times1.14$ \si{\mega \hertz}, with an estimated effective mass of approximately \SI{2}{\nano \gram}. The phononic crystal bandgap (Fig. \ref{fig:setup}(b))  isolates the mode from the substrate and flattens its profile near the clamping regions, suppressing strain and intrinsic losses (soft clamping) to achieve ultralow dissipation rates. Given the intrinsic mechanical quality factor of $Q\approx1.1\times10^8$, the corresponding mechanical damping rate $\Gamma_{\rm{m}}$ of this mode is $2\pi\times0.01$ \si{\hertz}. The ringdown measurement for characterizing the $Q$ factor is shown in Fig. \ref{fig:setup}(c). Such high $Q$ leads to a $Qf$ product of $1.3\times10^{14}$ corresponding to roughly 20 coherent oscillations, showing its potential for ground-state cooling at room temperature. The FP cavity used in the experiment has a finesse of approximately $2.7\times10^{4}$, composed of a curved mirror of $10$ ppm transmissivity and a flat mirror of $200$ ppm transmissivity, enabling effective enhancement of optomechanical coupling. The cavity linewidth $\kappa$ is $2\pi\times3.7$ \si{\mega \hertz} at a wavelength of \SI{1549.9}{\nano \meter}, which places the system in the near-sideband-resolved regime, enabling efficient DBC. However, due to the cavity birefringence, the resonance frequencies of the two polarization modes differ by $1.05\kappa$, which introduces additional optical loss and reduces the performance of the CFC. In addition, thermal motion of the cavity mirrors at room temperature is transduced by the cavity into frequency fluctuations, limiting the achievable cooling efficiency.

In the CFC experiment, a horizontally polarized beam is used as the probe field. After being reflected from the cavity, the beam polarization is rotated to vertical by a Faraday rotator. A small fraction of the reflected beam is tapped out to an avalanche photodetector (APD) for locking the laser to the cavity (see Supplement~1 for detailed setup). Around $2\%$ of the reflected beam is sent to a homodyne detector for optical readout. Although not required for the feedback protocol itself, this detection channel is used solely to characterize the cooling performance. The main reflected beam is combined with a strong auxiliary beam on a $90:10$ BS for the implementation of the displacement operation. This operation essentially rotates the carrier mode of the probe field in phase space, leaving the sidebands unrotated. By setting an appropriate displacement angle, the phase modulation imprinted by the mechanical motion is converted into amplitude modulation. The resulting field is then coupled to a fiber delay line, which implements the required temporal delay such that the displacement information maps onto the mechanical momentum quadrature. A small portion of the delayed beam is extracted to lock the relative phase of the displacement operation, and the remaining light is sent through a circulator and re-injected into the cavity to realize CFC. The overall efficiency of the feedback loop is estimated to be approximately $30\%$, taking into account the propagation losses in free-space and fiber components, beam tapping for locking and measurement, imperfect coupling to the fiber, and around $10\%$ loss in the displacement operation.

\section{Numerical Calculation}

In this section, we present numerical calculations based on the theoretical model to identify how the optimal parameters for CFC depend on the cavity properties. 

\begin{figure}
    \centering
    \includegraphics[width=1\linewidth]{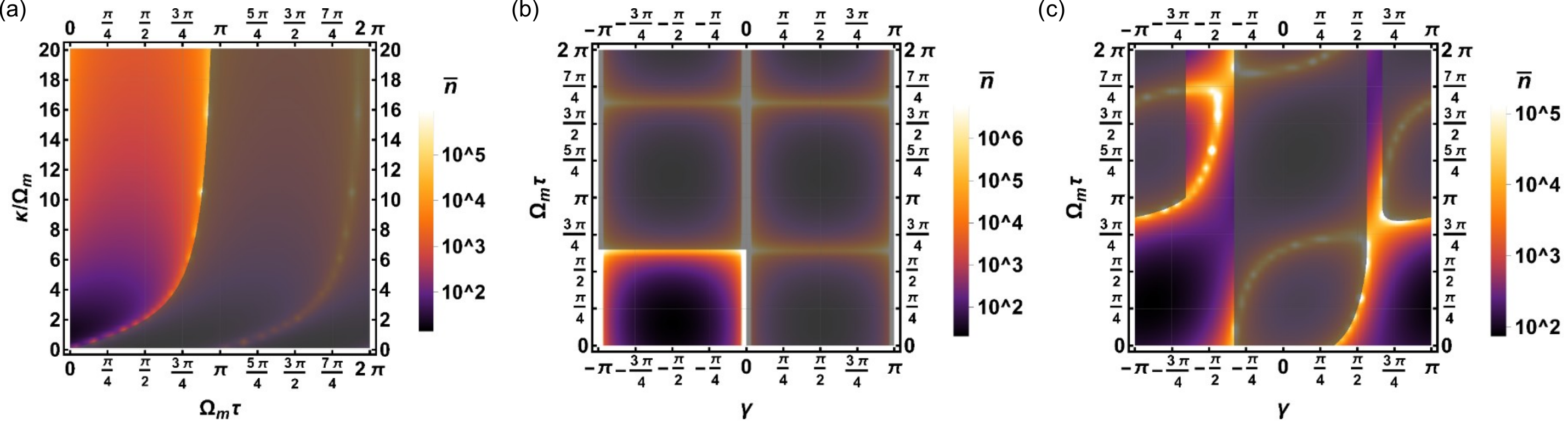}
    \captionsetup{width=1\linewidth}
    \caption{Numerical calculations of CFC. 
(a) Phonon occupation $\bar{n}$ versus the feedback delay $\Omega_{\rm{m}}\tau$ and the cavity linewidth $\kappa$ under the on-resonance condition, showing the influence of the cavity linewidth on the optimal delay.
(b) Phonon occupation $\bar{n}$ as a function of the feedback delay $\Omega_{\rm{m}}\tau$ and displacement angle $\gamma$ for $\kappa = 2\pi\times3.7$ \si{\mega \hertz}, with both optical modes on resonance. The optimal parameters are found near $\Omega_{\mathrm{m}}\tau \approx \pi/4$ and $\gamma \approx -\pi/2$.
(c) Phonon occupation $\bar{n}$ for the experimentally detunings $\Delta_{h} = -0.06\kappa$ and $\Delta_{v} = -1.11\kappa$. The optimal displacement angle shifts toward $-\pi$.
The unshaded region indicates the parameter space satisfying the sufficient stability criterion. Calculations above are performed using the following parameters, chosen to match our experimental conditions: $Q =1.1 \times 10^{8}$, $\Omega_{\rm{m}} = 2\pi\times1.14$ \si{\mega \hertz}, the escape efficiency $\eta_{\rm esc} = 0.68$, the input powers $P_{\rm{hin}} =$ \SI{50}{\micro \watt}, $P_{\rm{vin}} = $ \SI{270}{\micro \watt}, $g_{0,h} = 2\pi\times3.7$ \si{\hertz}, $g_{0,v} = 2\pi\times3.7$ \si{\hertz}, $\eta = 1$, and all additional noise terms are set to zero.
}
    \label{fig:simulation}
\end{figure}

Figure \ref{fig:simulation}(a) shows the influence of the cavity linewidth $\kappa$ on the optimal feedback delay (i.e., minimizes the phonon occupation) under the on-resonance condition. We find that when we are deeper into the fast-cavity regime ($\kappa>>\Omega_{\rm{m}}$), the optimal delay approaches $\pi/2$. As the cavity linewidth decreases, although the system is still not in the sideband-resolved regime, the cooling efficiency improves and the optimal delay shifts toward $\pi/4$. The displacement angle, defined as $\gamma = \phi-u + x$, is fixed to $-0.85\pi$ for Fig. \ref{fig:simulation}(a). For the on-resonance configuration, $\tau$ and $\gamma$ are independent parameters. The unshaded area corresponds to the region satisfying the sufficient criterion for the system stability (see Supplement~1 for details on the stability). When the system is stable, its response to perturbations decays over time, implying that all poles of the susceptibility function of the system have negative imaginary components. Note that points in grey-shaded region are not necessarily unstable since the adopted criterion is sufficient but not necessary. Points within the gray region may still be stable but require individual verification. Fig. \ref{fig:simulation}(b) shows the phonon occupation as a function of delay and displacement angle for a cavity linewidth fixed to our experimental value of $2\pi\times3.7$ \si{\mega \hertz}, with both optical modes on resonance. Within the stable region defined by the sufficient stability criterion, the optimal delay is near $\pi/4$, and the optimal displacement angle is close to $-\pi/2$. We also note that although a unique pair $(\gamma,\Omega_{\rm m} \tau)$ minimizes the phonon occupation, the cooling performance is relatively insensitive in a broad neighborhood around the optimum, indicating robustness of the protocol.

To investigate the effect of cavity detuning, Fig. \ref{fig:simulation}(c) presents the simulated phonon occupation as a function of $\Omega_{\rm{m}}\tau$ and $\gamma$ for detunings set to the experimental values $\Delta_h = -0.06\kappa$ and $\Delta_v = -1.11\kappa$, corresponding to the birefringence-induced splitting of the cavity modes. It is worth noting that when the cavity is exactly on resonance, the mechanical displacement is encoded purely on the phase quadrature of the light, and no additional quadrature rotation to the output optical mode is introduced by the cavity. By contrast, when the cavity is detuned, the mechanical displacement is encoded on both amplitude and phase quadratures of the output field. This affects the displacement operation, and the optimal conversion angle is influenced by the quadrature rotation induced by the cavity. In this case, these two parameters $(\gamma,\Omega_{\rm m} \tau)$ are no longer independent, and the optimal displacement angle shifts closer to $-\pi$ for the optimal delay, as shown in Fig. \ref{fig:simulation}(c).

\section{Experimental Results}

\begin{figure}
    \centering
    \includegraphics[width=1\linewidth]{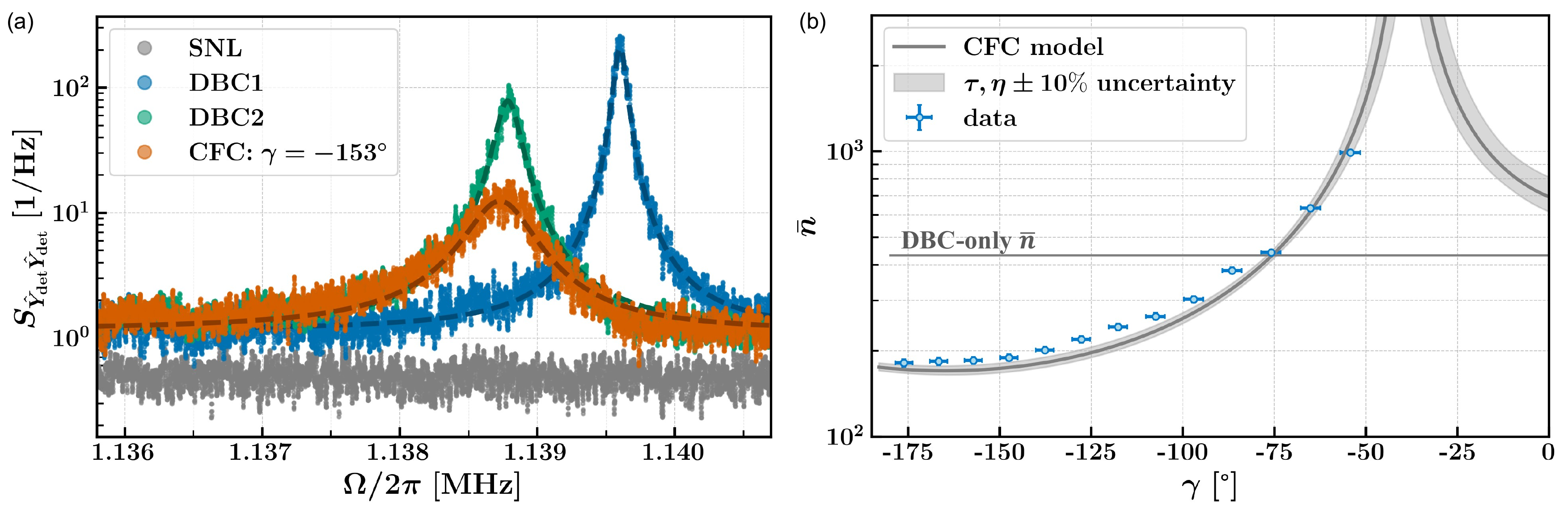}
    \captionsetup{width=1\linewidth}
    \caption{Experimental results of CFC.
(a) Phase-quadrature power spectral densities (PSDs) of the reflected probe beam measured by homodyne detection. The blue trace (DBC1) shows the DBC spectrum with only the probe beam (cooling beam blocked) at a detuning $\Delta_{h} = -0.06\kappa$ and $\sim$\SI{50}{\micro \watt} input probe power. The green trace (DBC2) shows the spectrum with both probe and cooling beams (feedback signal blocked), showing enhanced DBC effect with $\Delta_{v} = -1.11\kappa$ and $\sim$\SI{270}{\micro \watt}  input cooling power. The orange trace shows the spectrum with CFC with $\gamma = -153^\circ $, $\Omega_{\rm{m}}\tau = 0.24\pi$, and feedback efficiency $\eta = 0.3$. The grey trace shows the shot noise limit (SNL), normalized to 0.5. The dashed curves represent theoretical fits based on the model.
(b) Phonon occupation as a function of displacement angle $\gamma$. The grey curve shows the theoretical prediction. The shaded area represents the uncertainty estimated by including $10\%$ uncertainty in $\eta$ and $\tau$, which reflects our estimated experimental uncertainty for these quantities. Error bars in phonon number are obtained by error propagation, considering both the standard error of fitted parameters, and the standard error of integrating the areas used to obtain the phonon number for CFC. Error bars in $\gamma$ account for a $\pm 5^\circ$ uncertainty in the experimental lock phase and the uncertainty of the system parameters.
}
    \label{fig:result1}
\end{figure}

The cooling effects are illustrated in Fig. \ref{fig:result1}(a), which shows the phase quadrature PSDs of the reflected probe beam measured with a homodyne detector. The blue trace (DBC1) in Fig. \ref{fig:result1}(a) represents the case when only the probe beam is present, i.e., the cooling beam is blocked. In this configuration, the probe beam is slightly red detuned, leading to a shifted and broadened mechanical spectrum due to DBC. By fitting this spectrum with our model (dashed line), the detuning level $\Delta_{h} = -0.06\kappa$, the coupling strength $g_{0,h} = 2\pi\times3.7$ \si{\hertz}, and the additional frequency fluctuation term of the probe beam $S_{\delta\Delta_{h}\delta\Delta_{h}}$ can be extracted. Then, the cooling beam is unblocked while the feedback signal remains blocked, i.e., the added cooling effect only comes from the DBC introduced by the displacement operation's auxiliary beam (DBC2). This case is shown as the green trace. By fitting the model to this spectrum (dashed line), $g_{0,v} = 2\pi\times 3.7$ \si{\hertz} and the additional frequency fluctuation term of the cooling beam $S_{\delta\Delta_{v}\delta\Delta_{v}}$ are extracted. By reconstructing $S_{\hat{Q}\hat{Q}}$ with all system parameters and numerically integrating it, the phonon number under DBC alone is determined to be $444\pm 6$ for a probe power of \SI{50}{\micro \watt}  and a cooling power of \SI{270}{\micro \watt}, demonstrating effective DBC with this relatively narrow cavity linewidth. The contribution to the phonon number from additional noise sources is about $136$, where the extra noise arise from mirror vibrations, laser-lock instability, laser phase and amplitude noise, and thermal intermodulation noise. This estimate is obtained by evaluating the fitted model with the technical-noise terms set to zero and attributing the difference in $\bar n$ to excess noise. Finally, we unblock the feedback signal, as shown by the orange trace (CFC) in Fig. \ref{fig:result1}(a), where a clear CFC effect is observed leading to a final thermal occupation of $\bar{n} = 185 \pm 5$ phonons.
This is measured with delay $\Omega_{\rm m} \tau = 0.24\pi$, and a displacement angle $\gamma = -0.85 \pi$ (see Supplement~1 for the relation between the interference angle $\theta$ and the displacement angle $\gamma$ defined in the model). The CFC fit (dashed line) is obtained by setting $\gamma$ as the only free parameter. In addition, Supplement~1 presents an example fit based on a more complete model for DBC, in which the noise sources previously grouped into the single effective term $S_{\delta\Delta\delta\Delta}$ are modeled separately.  
Including separate laser amplitude and phase noise terms reproduces the observed asymmetry in the tails of the detected spectrum, which cannot be captured by a single effective noise term.

By varying the relative phase between the signal and auxiliary beams in the displacement operation, we investigate the effect of the displacement angle in the thermal occupation number $\bar{n}$ while keeping the detunings, optical powers, and delay fixed. The resulting dependence of the phonon number on the displacement angle, one of the key parameters of CFC, is shown in Fig. \ref{fig:result1}(b). The grey curve represents the theoretical prediction. The experimental phonon number is obtained by comparing the areas of the spectra with only DBC and with CFC (see Supplement~1 for more details), which serves as an independent method to validate the phonon-number estimation under CFC. The experimental data points show good agreement with the model. Depending on the displacement angle, the feedback force can either cool or heat the resonator, and the optimal CFC is achieved when the displacement angle approaches $-\pi$.

\begin{figure}
    \centering
    \includegraphics[width=1\linewidth]{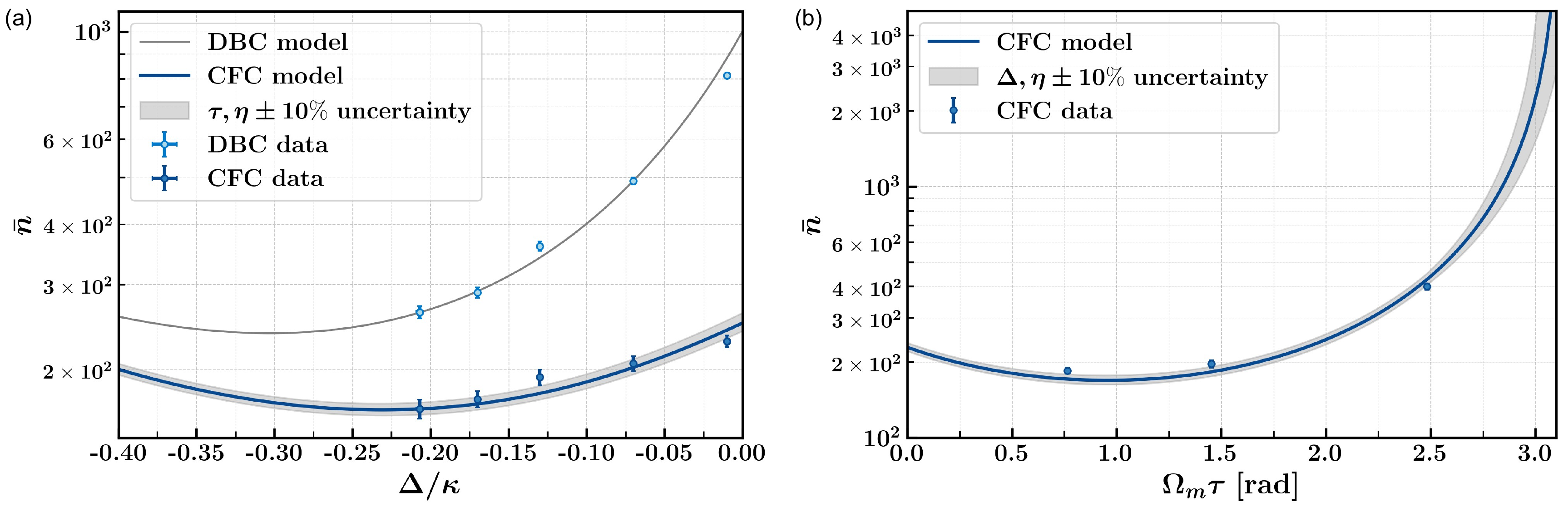}
    \captionsetup{width=1\linewidth}
    \caption{CFC as a function of detuning and delay. (a) Phonon occupation as a function of the probe detuning $\Delta_h/\kappa$. The light grey curve shows the theoretical prediction for DBC, while the dark blue curve includes both DBC and CFC. The shaded grey area represents the uncertainty in phonon number estimated by including $10\%$ variation in $\eta$ and $\Omega_{\rm m}\tau$. Light and dark blue circles denote the experimental data for DBC and CFC, respectively. (b) Phonon occupation versus the delay $\Omega_{\rm m}\tau$ with $\gamma = -0.85\pi$ and $\Delta_h/\kappa = -0.06$. The dark blue curve represents the theoretical prediction, and the shaded region shows the uncertainty considering $10\%$ fluctuations in $\Delta_{h}$ and $\eta$. Experimental data points (blue circles) agree well with the theoretical curve. The error bars represent the uncertainty propagated from the fitted parameters and integration area used to extract the phonon number.
}
    \label{fig:result2}
\end{figure}

We further investigate the CFC performance by scanning the detuning and the delay. Fig. \ref{fig:result2}(a) shows the phonon occupation versus the detuning of the probe beam while keeping $\gamma$ and $\Omega_{\rm m}\tau$ fixed. The light grey curve corresponds to the theoretical prediction with only DBC, while the dark blue curve includes both DBC and CFC effects. For DBC, the phonon number decreases with increasing red detuning of the probe beam and reaches a minimum at $\Delta_{h}/\kappa \sim -0.3$, close to the mechanical frequency $\Omega_{\rm{m}}$. The phonon number achieved with CFC surpasses that obtained from optimal DBC with $\Delta_{h} = -\Omega_{\rm{m}}$ under fixed power conditions. 
At higher powers required for deeper cooling, technical noise becomes increasingly relevant and can offset cooling gains.
CFC, in this case, shows an advantage in further cooling the resonator beyond the DBC limit. The experimental data points show agreement with the theoretical predictions for both DBC and CFC. The lowest phonon number obtained with CFC is $166\pm7$, with experimental parameters: $\Delta_{h}/\kappa = -0.21$, $\gamma = -0.85\pi$, and $\Omega_{\rm m}\tau = 0.24\pi$.
Fig. \ref{fig:result2}(b) shows the phonon number as a function of the delay $\Omega_{\rm m}\tau$, which serves as another crucial parameter for CFC. The dark blue curve shows the theoretical prediction obtained with fixed parameters of $\gamma = -0.85\pi$ and $\Delta_{h}/\kappa = -0.06$. The data points, measured with $\Omega_{\rm m}\tau = 0.24\pi$, $0.46\pi$, and $0.79\pi$, show agreement with the theoretical prediction.

\section{Discussion}

\begin{figure}[ht]
    \centering
    \includegraphics[width=1\linewidth]{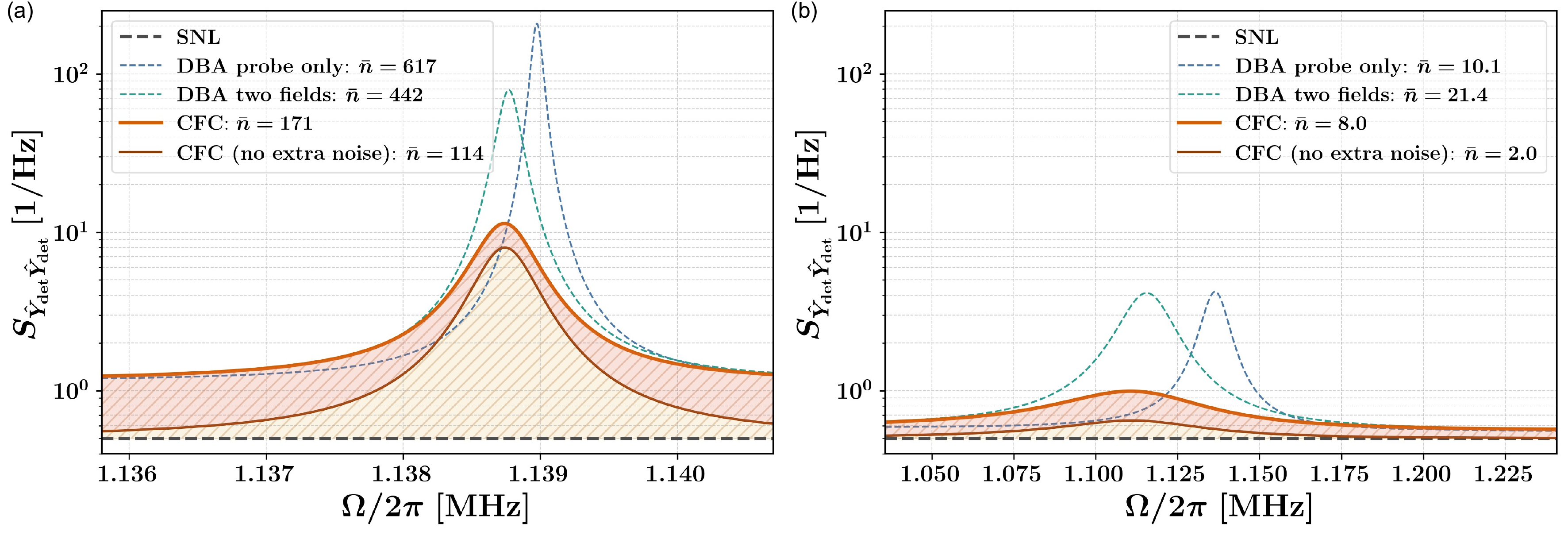}
    \caption{Predicted performance gains from realistic improvements in $g_0$ and frequency-noise suppression. Simulated detected spectra $S_{Y_{\rm det}Y_{\rm det}}(\Omega)$ for (a) the current experimental parameter set and (b) an upgraded scenario with a $10$-fold increase in the single-photon coupling $g_0$ and a \SI{10}{\decibel} reduction of cavity frequency noise. The dashed curves show DBC with the probe field only and with both probe and cooling fields, while the solid curves show CFC. The SNL is indicated for reference. The curve labeled ``CFC (no extra noise)'' is obtained by setting the technical-noise terms (cavity-frequency noise and laser amplitude/phase noise) to zero, thereby illustrating the quantum-noise-limited performance.}
    \label{fig:CFC_simulation}
\end{figure}

A cornerstone of our framework is that it goes beyond idealized CFC models by explicitly accounting for experimentally relevant technical noise sources. These external noise terms drive intracavity fluctuations that are subsequently processed by the feedback loop and can therefore become performance-limiting in the same regimes where CFC would otherwise provide the largest cooling benefit. This is visible in Fig.~\ref{fig:CFC_simulation}(a), where the experimentally realistic CFC prediction ($\bar{n}=171$), using the same parameters as the CFC trace in Fig. 4(a), remains above the idealized ``CFC (no extra noise)'' limit ($\bar{n}=114$), demonstrating that technical noise already accounts for a substantial fraction of the residual occupation. 
 
Figure~\ref{fig:CFC_simulation}(b) illustrates how realistic improvements in device parameters and noise mitigation can yield major performance gains. Assuming a conservative $10$-fold increase in $g_0$ together with a $10$-fold reduction of cavity-frequency noise, the model predicts a reduction to $\bar{n}=8.0$ for CFC, while the quantum-noise-limited prediction approaches $\bar{n}=2.0$. In this upgraded regime, the remaining gap between the realistic CFC trace and the ``no extra noise'' trace directly quantifies how strongly external noise sources set the cooling floor, and thus pinpoints where further experimental effort should be focused.
 
Importantly, the improvement assumptions used here are realistic and strongly motivated by the literature. For example, Huang \textit{et al.}\ demonstrated that phononic-structure engineering of cavity mirrors can suppress cavity-frequency noise by more than \SI{20}{\decibel}, i.e.\ well beyond the conservative \SI{10}{\decibel} reduction considered here (see Ref.~\cite{huangRoomtemperatureQuantumOptomechanics2024a}). Likewise, a $10$-fold increase in $g_0$ would bring our coupling strength into the range of standard optomechanical platforms. For reference, Ernzer \textit{et al.}\ reported $g_0/2\pi = $ \SI{160}{\hertz}. Taken together, these simulations emphasize that the same model that reproduces the current experiment can also serve as a quantitative design tool: by selectively enabling/disabling specific noise contributions and varying experimentally achievable parameters (e.g., $g_0$, feedback-loop loss, and cavity-frequency noise), one can plan realistic upgrade paths toward targeted goals such as $\bar{n}\lesssim 1$ (ground-state cooling) without cryogenics. 

\section{Conclusion}
In conclusion, we have experimentally demonstrated CFC of a high-Q density phononic crystal membrane at room temperature, and developed a theoretical model that incorporates experimentally relevant loss and technical noise terms. We simulated how cavity parameters influence the optimal conditions for CFC in the near-sideband-resolved regime, revealing deviations from the fast-cavity limit. Experimentally, we characterized how the cooling performance depends on displacement angle and implemented CFC across a range of detunings and optical delays, finding good agreement with the model. By combining CFC with strong DBC enabled by the near-sideband-resolved cavity, we achieved a minimum phonon occupation of  $166\pm7$.

To further enhance the system performance, several avenues can be pursued: increasing the optomechanical coupling ($g_0$) to increase feedback gain, reducing in-loop optical losses to improve loop efficiency, and suppressing cavity-mirror vibrations (e.g., via phononic-bandgap engineering) to reduce excess frequency noise at room temperature. Using the validated noise model as a design tool, we project that realistic improvements such as a $10$-fold increase in $g_0$ and a factor of $10$ reduction in cavity-frequency noise would enable CFC to reach $\bar n \sim 8$ (and $\bar n \sim 2$ in the quantum-noise-limited case) at room temperature, delineating a feasible route toward ground-state operation without cryogenics.

These results position coherent optical feedback as a practical control approach for optomechanical platforms. The agreement with a realistic model provides a transferable framework for designing delayed coherent feedback protocols in noisy, non-ideal cavities. This advances room-temperature coherent optical control of macroscopic mechanical systems and supports progress toward ground-state cooling.

\section*{Funding}
This work was funded by the Danish National Research Foundation, Center for Macroscopic Quantum States (bigQ, DNRF0142).


\bibliography{refs}

\clearpage 

\appendix

\setcounter{equation}{0}
\setcounter{figure}{0}
\setcounter{table}{0}

\renewcommand{\theequation}{S\arabic{equation}}
\renewcommand{\thefigure}{S\arabic{figure}}
\renewcommand{\thetable}{S\arabic{table}}

\renewcommand{\theHequation}{S\arabic{equation}}
\renewcommand{\theHfigure}{S\arabic{figure}}
\renewcommand{\theHtable}{S\arabic{table}}

\title{Coherent Feedback Cooling of an Ultracoherent Phononic-Crystal Membrane at Room Temperature: supplemental document}
\author{} 

\section{Theoretical model}
\subsection{Solving the Langevin equations}

The Langevin equations describing the system are given by

\begin{align}
	\dot{\hat{Q}} =& \Omega_{\rm m} \hat{P}, 
	\\
	\dot{\hat{P}} =& -\Omega_{\rm m}\hat{Q}-\Gamma_{\rm m} \hat{P} -\sqrt{2}g_{0,h}\hat{h}^{\dagger}\hat{h}-\sqrt{2}g_{0,v}\hat{v}^{\dagger}\hat{v}+ \sqrt{2\Gamma_{\rm m}}\hat{P}_{\text{in}},
	\\
	\dot{\hat{h}} =& -\frac{\kappa}{2}\hat{h} + i (\Delta_{h}+\delta\Delta_{h})\hat{h}-i\sqrt{2}g_{0,h}\hat{Q}\hat{h} +\sqrt{\kappa^{\text{in}}}\hat{h}_{\text{in}}+\sqrt{\kappa-\kappa^{\rm in}}\hat{h}_{\text{loss}},
	\\
	\begin{split}
	\dot{\hat{v}} =& -\frac{\kappa}{2}\hat{v}+i (\Delta_{v}+\delta\Delta_{v})\hat{v}-i\sqrt{2}g_{0,v}\hat{Q}\hat{v} +\sqrt{\kappa^{\text{in}}\eta}e^{-i\phi}T_{\tau}\left(
	\hat{h}_{\text{in}}-\sqrt{\kappa^{\text{in}}}\hat{h}
	\right)\\
	&+\sqrt{\kappa^{\text{in}}}\delta +\sqrt{\kappa-\eta\kappa^{\rm{in}}}\hat{v}_{\rm{v}}.
	\end{split}
\end{align}
\noindent where the term $\sqrt{\kappa-\eta\kappa^{\rm{in}}}\hat{v}_{\rm{v}}$ represents a combination of two vacuum noise sources, satisfying $\sqrt{\kappa-\kappa^{\mathrm{in}}}\,\hat{v}_{\mathrm{loss}}
+ \sqrt{\kappa^{\mathrm{in}}(1-\eta)}\,\hat{h}_{v}
= \sqrt{\kappa-\kappa^{\mathrm{in}} + \kappa^{\mathrm{in}}(1-\eta)}\,\hat{v}_{\mathrm{v}}
= \sqrt{\kappa - \eta \kappa^{\mathrm{in}}}\,\hat{v}_{\mathrm{v}}$.

\noindent To linearize above Langevin equations,  we first derive the steady-state solutions. From the equations for the expectation values

\begin{align}
	\dot{\braket{Q}} =& \Omega_{\rm m} \braket{P},
	\\
	\dot{\braket{P}} =&-\Omega_{\rm m} \braket{Q}-\Gamma_{\rm m} \braket{P}-\sqrt{2}g_{0,h}\braket{h^{\dagger}h}-\sqrt{2}g_{0,v}\braket{v^{\dagger}v},\\
	\dot{\braket{h}} =& -\frac{\kappa}{2}\braket{h} + i \Delta_{h}\braket{h}+i \braket{\delta\Delta_{h}h}-i\sqrt{2}g_{0,h}\braket{Qh} +\sqrt{\kappa^{\text{in}}}\braket{h_{\text{in}}},
	\\
	\dot{\braket{v}} =& -\frac{\kappa}{2}\braket{v} + i \Delta_{v}\braket{v}+i \braket{\delta\Delta_{v}v}-i\sqrt{2}g_{0,v}\braket{Qv} \\ \nonumber
&+\sqrt{\kappa^{\text{in}} \eta}e^{-i\phi}\left(
	\braket{T_{\tau}h_{\text{in}}}-\sqrt{\kappa^{\text{in}}}\braket{T_{\tau}h}
	\right)+\sqrt{\kappa^{\text{in}}}\delta,
\end{align}

\noindent  with $\braket{P_{\rm in}} = \braket{\hat{h}_{\rm loss}} = \braket{\hat{v}_{\rm v}} = 0$, the steady-state condition $\dot{\braket{Q}} = \dot{\braket{P}} = \dot{\braket{h}} = \dot{\braket{v}} = 0$ leads to:

\begin{align}
0 &= \Omega_{\rm m} \braket{P},\\
0 &=-\Omega_{\rm m}\braket{Q}-\Gamma_{\rm m} \braket{P} - \sqrt{2}g_{0,h}\lvert \braket{h}\rvert^2-\sqrt{2}g_{0,v}\lvert \braket{v}\rvert^2, \\
0  &= -\frac{\kappa}{2}\braket{h}+i \Delta_h^{\text{eff}}\braket{h} +\sqrt{\kappa^{\text{in}}}\braket{h_{\text{in}}}, \\
0 &= -\frac{\kappa}{2}\braket{v}+i \Delta_{v}^{\text{eff}}\braket{v} +\sqrt{\kappa^{\text{in}} \eta}e^{-i\phi}\left[
	\braket{h_{\text{in}}}-\sqrt{\kappa^{\text{in}}}\braket{h}\right]+\sqrt{\kappa^{\text{in}}}\delta,
\end{align}

\noindent where $\braket{T_{\tau}h_{\text{in}}} = \braket{h_{\rm in}}$ and $\braket{T_{\tau}h} = \braket{h}$. The effective detunings $\Delta_{h}^{\text{eff}}$ and $\Delta_{v}^{\text{eff}}$ are defined as:

\begin{align}
	\Delta_{h}^{\text{eff}} := \Delta_{h}+\braket{\delta\Delta_{h}} - \sqrt{2}g_{0,h}\braket{Q}, \\
	\Delta_{v}^{\text{eff}} := \Delta_{v}+\braket{\delta\Delta_{v}} - \sqrt{2}g_{0,v}\braket{Q}.	
\end{align}

\noindent Here, second order moments are neglected by applying $\braket{AB} = \braket{(\braket{A}+\delta A)(\braket{B}+\delta B)}\approx\braket{A}\braket{B}$.  
\noindent The resulting steady-state solutions are given by:

\begin{align}
	\braket{h} &= \frac{\sqrt{\kappa^{\text{in}}}}{-i\Delta_{h}^{\text{eff}} +\kappa/2}\braket{h_{\text{in}}},
	\\
	\braket{v} &= \frac{\sqrt{\kappa^{\text{in}}}}{-i\Delta_{v}^{\text{eff}}+\kappa/2}\left[\delta + \sqrt{\eta}e^{-i\phi}\left(1-\frac{\kappa^{\text{in}}}{-i\Delta_{h}^{\text{eff}} +\kappa/2}\right)\braket{h_{\text{in}}}\right],
	\\
	\braket{Q} &= \frac{\sqrt{2}}{\Omega_{\rm m}}\left(g_{0,h}\lvert\braket{h}\rvert^2 + g_{0,v}\lvert\braket{v}\rvert^2 \right),
	\\
	\braket{P} &= 0.
\end{align}

\noindent With the steady-state values determined, we now consider the dynamics of fluctuations terms. With the definitions $g_{h}:= g_{0,h} \braket{h} = |g_{h}|e^{iu}$, $g_{v}:= g_{0,v} \braket{v} = |g_{v}|e^{ix}$ and the linearization $\delta(\hat{A}\hat{B}) = \hat{A}\hat{B}-\braket{AB}\approx\delta\hat{A}\braket{B}+\braket{A}\delta\hat{B}$, the equations for the fluctuation terms can then be linearized as

\begin{align}
	\delta\dot{\hat{Q}} =& \Omega_{\rm m} \delta\hat{P}, 
	\\
	\delta\dot{\hat{P}} =& -\Omega_{\rm m} \delta\hat{Q} - \Gamma_{\rm m} \delta\hat{P} - \sqrt{2}g_{h}(\delta\hat{h} + \delta\hat{h}^{\dagger}) - \sqrt{2}|g_{v}|(e^{-ix}\delta\hat{v} + e^{ix}\delta\hat{v}^{\dagger}) + \sqrt{2\Gamma_{\rm m}}\hat{P}_{\text{in}},
	 \\
	\delta\dot{\hat{h}} =& - \frac{\kappa}{2}\delta\hat{h} +i\Delta_{h}^{\rm eff}\delta\hat{h}+i\braket{h}\delta\Delta_{h}-i\sqrt{2}g_{h}\delta\hat{Q} + \sqrt{\kappa^{\text{in}}}\delta\hat{h}_{\text{in}} + \sqrt{\kappa-\kappa^{\text{in}}}\delta\hat{h}_{\text{loss}}, 
	\\
	\delta\dot{\hat{v}} =& - \frac{\kappa}{2}\delta\hat{v}+i\Delta_{v}^{\rm eff}\delta\hat{v}+i\braket{v}\delta\Delta_{v}-i \sqrt{2}g_{v}\delta\hat{Q}  \\  \nonumber
	&+ \sqrt{\kappa^{\text{in}} \eta}e^{-i\phi}T_{\tau}(\delta\hat{h}_{\text{in}} - \sqrt{\kappa^{\text{in}}}\delta\hat{h}) +\sqrt{\kappa- \eta\kappa^{\rm{in}}}\delta\hat{v}_{\rm{v}}.
\end{align}

\noindent To simplify the subsequent analysis, we introduce phase-rotated field operators $\hat{h}^{u} = e^{-iu} \hat{h}$ and $\hat{v}^{x} = e^{-ix} \hat{v}$, such that all coupling strength $g_{h}$ and $g_{v}$ can be substituted with their absolute values $|g_{h}|$ and $|g_{v}|$. The linearized Langevin equations can be rewritten as 

\begin{align}
	\delta\dot{\hat{Q}} =& \Omega_{\rm m} \delta\hat{P}, 
	\\
	\delta\dot{\hat{P}} =& -\Omega_{\rm m} \delta\hat{Q} - \Gamma_{\rm m} \delta\hat{P} - \sqrt{2}|g_{h}|(\delta\hat{h}^{u} + (\delta\hat{h}^{u})^{\dagger}) - \sqrt{2}|g_{v}|(\delta\hat{v}^{x} + (\delta\hat{v}^{x})^{\dagger}) + \sqrt{2\Gamma_{\rm m}}\hat{P}_{\text{in}},
	 \\
	\delta\dot{\hat{h}}^{u} =& - \frac{\kappa}{2}\delta\hat{h}^{u} +i\Delta_{h}^{\rm eff}\delta\hat{h}^{u}+i|\alpha_{h,\rm{cav}}|\delta\Delta_{h}-i\sqrt{2}|g_{h}|\delta\hat{Q} + \sqrt{\kappa^{\text{in}}}\delta\hat{h}^{u}_{\text{in}} + \sqrt{\kappa-\kappa^{\text{in}}}\delta\hat{h}_{\text{loss}}^{u}, 
	\\
	\delta\dot{\hat{v}}^{x} =& - \frac{\kappa}{2}\delta\hat{v}^{x}+i\Delta_{v}^{\rm eff}\delta\hat{v}^{x}+i|\alpha_{v,\rm{cav}}|\delta\Delta_{v}-i \sqrt{2}|g_{v}|\delta\hat{Q}  \\  \nonumber
	&+ \sqrt{\kappa^{\text{in}} \eta}e^{-i(\phi-u+x)}T_{\tau}(\delta\hat{h}^{u}_{\text{in}} - \sqrt{\kappa^{\text{in}}}\delta\hat{h}^{u}) +\sqrt{\kappa- \eta\kappa^{\rm{in}}}\delta\hat{v}_{\rm{v}}^{x}.
\end{align}

The equations are then solved in Fourier space with quadratures:

\begin{align}
-i\omega\, \hat{Q}(\omega) &= \Omega_{\rm m}\hat{P}(\omega), \\[0pt]
-i\omega\, \hat{P}(\omega) &= -\Omega_{\rm m}\hat{Q}(\omega) - \Gamma_{\rm m} \hat{P}(\omega) - 2 |g_{h}|\, \hat{X}_h(\omega) - 2 |g_{v}|\, \hat{X}_v(\omega)
 + \sqrt{2\Gamma_{\rm m}}\, \hat{P}_{\mathrm{in}}(\omega), \\[2pt]
-i\omega\, \hat{X}_h(\omega) &= -\frac{\kappa}{2} \hat{X}_h(\omega)
 - \Delta_{h}^{\rm eff}\, \hat{Y}_h(\omega)
 + \sqrt{\kappa^{\mathrm{in}}}\, \hat{X}_h^{\mathrm{in}}(\omega)
 + \sqrt{\kappa-\kappa^{\mathrm{in}}}\, \hat{X}_h^{\mathrm{loss}}(\omega), \\[2pt]
-i\omega\, \hat{Y}_h(\omega) &= -2 |g_{h}|\, \hat{Q}(\omega)
 + \Delta_{h}^{\rm eff}\, \hat{X}_h(\omega)
 - \sqrt{2}\, |\alpha_{h,\mathrm{cav}}|\, \delta\Delta_h(\omega)
 - \frac{\kappa}{2}\, \hat{Y}_h(\omega) \nonumber\\
&\quad + \sqrt{\kappa^{\mathrm{in}}}\, \hat{Y}_h^{\mathrm{in}}(\omega)
 + \sqrt{\kappa-\kappa^{\mathrm{in}}}\, \hat{Y}_h^{\mathrm{loss}}(\omega), \\[2pt]
-i\omega\, \hat{X}_v(\omega) &= -\frac{\kappa}{2} \hat{X}_v(\omega)
 - \Delta_{v}^{\rm eff}\, \hat{Y}_v(\omega)
 + \sqrt{\kappa - \eta\,\kappa^{\mathrm{in}}}\, \hat{X}_v^{\mathrm{loss}}(\omega) \nonumber\\
&\quad + \sqrt{\eta\,\kappa^{\mathrm{in}}}\, e^{i\omega\tau}
\!\left( \hat{X}_h^{\mathrm{in}}(\omega)\cos\gamma
      + \hat{Y}_h^{\mathrm{in}}(\omega)\sin\gamma \right)
  \nonumber\\
  &\quad - \sqrt{\eta}\,\kappa^{\mathrm{in}}\, e^{i\omega\tau}
\!\left( \hat{X}_h(\omega)\cos\gamma
      + \hat{Y}_h(\omega)\sin\gamma \right), \\[2pt]
-i\omega\, \hat{Y}_v(\omega) &= -2 |g_{v}|\, \hat{Q}(\omega)
 + \Delta_{v}^{\rm eff}\, \hat{X}_v(\omega)
 - \sqrt{2}\, |\alpha_{v,\mathrm{cav}}|\, \delta\Delta_v(\omega)
 - \frac{\kappa}{2}\, \hat{Y}_v(\omega)\nonumber\\
 &\quad+ \sqrt{\kappa - \eta\,\kappa^{\mathrm{in}}}\, \hat{Y}_v^{\mathrm{loss}}(\omega)
 + \sqrt{\eta\,\kappa^{\mathrm{in}}}\, e^{i\omega\tau}
\!\left( -\hat{X}_h^{\mathrm{in}}(\omega)\sin\gamma
      + \hat{Y}_h^{\mathrm{in}}(\omega)\cos\gamma \right)
  \nonumber\\
  &\quad - \sqrt{\eta}\,\kappa^{\mathrm{in}}\, e^{i\omega\tau}
\!\left( -\hat{X}_h(\omega)\sin\gamma
      + \hat{Y}_h(\omega)\cos\gamma \right),
\end{align}

\noindent where $\gamma = \phi - u + x$ represents the total displacement angle. For notational simplicity, the superscripts $x$ and $u$ are omitted in the following expressions. In practice, we solve for $\hat{X}^{u}_{h}(\omega)$, $\hat{Y}^{u}_{h}(\omega)$, $\hat{X}^{x}_{v}(\omega)$ and $\hat{Y}^{x}_{v}(\omega)$, which can be seen as taking the intracavity field as the phase reference. These Langevin equations can be written in matrix form as

\begin{equation}
-i\omega \hat{\mathbf{X}}(\omega) = \mathbf{A}(\omega)\hat{\mathbf{X}}(\omega) + \hat{\mathbf{b}}(\omega),
\end{equation}
with
\begin{equation}
\hat{\mathbf{X}}(\omega) =
\begin{bmatrix}
\hat{Q}(\omega) \\[0pt]
\hat{P}(\omega) \\[0pt]
\hat{X}_h(\omega) \\[0pt]
\hat{Y}_h(\omega) \\[0pt]
\hat{X}_v(\omega) \\[0pt]
\hat{Y}_v(\omega)
\end{bmatrix},
\end{equation}

\begin{equation}
\mathbf{A}(\omega) =
\begin{bmatrix}
0 & \Omega_{\rm m} & 0 & 0 & 0 & 0 \\[0pt]
-\Omega_{\rm m} & -\Gamma_{\rm m} & -2|g_h| & 0 & -2|g_v| & 0 \\[0pt]
0 & 0 & -\frac{\kappa}{2} & -\Delta_{h}^{\rm eff} & 0 & 0 \\[0pt]
-2|g_h| & 0 & \Delta_{h}^{\rm eff} & -\frac{\kappa}{2} & 0 & 0 \\[0pt]
0 & 0 &
-\sqrt{\eta}\,\kappa^{\mathrm{in}}\, e^{i\omega\tau}\cos\gamma &
-\sqrt{\eta}\,\kappa^{\mathrm{in}}\, e^{i\omega\tau}\sin\gamma &
-\frac{\kappa}{2} & -\Delta_{v}^{\rm eff} \\[0pt]
-2|g_v| & 0 &
\sqrt{\eta}\,\kappa^{\mathrm{in}}\, e^{i\omega\tau}\sin\gamma &
-\sqrt{\eta}\,\kappa^{\mathrm{in}}\, e^{i\omega\tau}\cos\gamma &
\Delta_{v}^{\rm eff} & -\frac{\kappa}{2}
\end{bmatrix},
\end{equation}

\begin{small}
\begin{equation}
\hat{\mathbf{b}}(\omega) =
\begin{bmatrix}
0 \\[0pt]
\sqrt{2\Gamma_{\rm m}}\,\hat{P}_{\mathrm{in}}(\omega) \\[0pt]
\sqrt{\kappa^{\mathrm{in}}}\,\hat{X}_h^{\mathrm{in}}(\omega)
+ \sqrt{\kappa-\kappa^{\mathrm{in}}}\,\hat{X}_h^{\mathrm{loss}}(\omega) \\[0pt]
\sqrt{\kappa^{\mathrm{in}}}\,\hat{Y}_h^{\mathrm{in}}(\omega)
+ \sqrt{\kappa-\kappa^{\mathrm{in}}}\,\hat{Y}_h^{\mathrm{loss}}(\omega)
\;\;-\;\sqrt{2}\,|\alpha_{h,\mathrm{cav}}|\,\delta\Delta_h(\omega) \\[0pt]
\sqrt{\kappa - \eta\,\kappa^{\mathrm{in}}}\,\hat{X}_v^{\mathrm{loss}}(\omega)
+ \sqrt{\eta\,\kappa^{\mathrm{in}}}\, e^{i\omega\tau}
\!\left(\hat{X}_h^{\mathrm{in}}(\omega)\cos\gamma
      + \hat{Y}_h^{\mathrm{in}}(\omega)\sin\gamma \right) \\[0pt]
\sqrt{\kappa - \eta\,\kappa^{\mathrm{in}}}\,\hat{Y}_v^{\mathrm{loss}}(\omega)
+ \sqrt{\eta\,\kappa^{\mathrm{in}}}\, e^{i\omega\tau}
\!\left(-\hat{X}_h^{\mathrm{in}}(\omega)\sin\gamma
      + \hat{Y}_h^{\mathrm{in}}(\omega)\cos\gamma \right)
\;\;-\;\sqrt{2}\,|\alpha_{v,\mathrm{cav}}|\,\delta\Delta_v(\omega)
\end{bmatrix}.
\end{equation}
\end{small}

\noindent The solution can be expressed as:
\begin{equation}
\hat{\mathbf{X}}(\omega)
= -(\mathbf{A}(\omega) + i\omega \mathbf{I})^{-1}\hat{\mathbf{b}}(\omega),
\end{equation}

\noindent where $\mathbf{I}$ is the identity matrix.

\noindent Especially, the solution of mechanical displacement $\hat{Q}(\omega)$ can be expressed as:
\begin{equation}
\hat{Q}(\omega) = \chi_{\mathrm{cf}}(\omega)\,\hat{\xi}(\omega)\mathbf{T}(\omega),
\end{equation}

\noindent where $\chi_{\mathrm{cf}}(\omega)$ denotes the susceptibility function of the coherent feedback system, $\hat{\xi}$ is the vector of the input quadratures (The full expression of matrix $\mathbf{T}(\omega)$ is lengthy and therefore not shown here):

\begin{align}
\left[\chi_{\mathrm{cf}}\right]^{-1} = &i (16 \Delta_{v}^{\rm eff} \left(4 {\Delta_{h}^{\rm eff}}^2+(\kappa -2 i \omega )^2\right) |g_{v}|^2+\left(4 {\Delta_{v}^{\rm eff}}^2+(\kappa -2 i \omega )^2\right) \nonumber \\
&\left(16 |g_{h}|^2 \Delta_{h}^{\rm eff}-\left(4 {\Delta_{h}^{\rm eff}}^2+(\kappa -2 i \omega )^2\right) \left(\omega ^2+i \Gamma_{\rm m} \omega -1\right)\right))\nonumber \\
&-32 e^{i \tau  \omega } |g_{h}||g_{v}|(\Delta_{h}^{\rm eff}+\Delta_{v}^{\rm eff}) \sqrt{\eta } \kappa^{\rm in} (i \kappa +2 \omega ) \cos \gamma \nonumber \\
&+16 e^{i \tau  \omega }|g_{h}||g_{v}|\kappa^{\rm in} \left(i (\kappa -2 i \omega )^2-4 i \Delta_{h}^{\rm eff} \Delta_{v}^{\rm eff}\right) \sin \gamma \sqrt{\eta },
\end{align}

\begin{small}
\begin{equation}
\hat{\xi}(\omega) := 
\begin{bmatrix}
\hat{P}_{\mathrm{in}}(\omega) &
\hat{X}^{\mathrm{in}}_{h}(\omega) &
\hat{Y}^{\mathrm{in}}_{h}(\omega) &
\hat{X}^{\mathrm{loss}}_{h}(\omega) &
\hat{Y}^{\mathrm{loss}}_{h}(\omega) &
\hat{X}^{\mathrm{loss}}_{v}(\omega) &
\hat{Y}^{\mathrm{loss}}_{v}(\omega) &
\delta\Delta_{h}(\omega) &
\delta\Delta_{v}(\omega)
\end{bmatrix}.
\end{equation}
\end{small}

\noindent The stability of the system can be evaluated from the poles of $\chi_{\mathrm{cf}}(\omega)$. The system is stable as long as all poles $\omega_{p}$ defined by $[\chi(\omega_{p})]^{-1} = 0$ remain in the lower half of the complex frequency plane (Im$(\omega_{p})<0$), ensuring that any oscillations decay over time rather than grow \cite{Luiz24}.

\noindent The power spectral density (PSD) of the mechanical displacement can be expressed as:
\begin{equation}
\begin{aligned}
S_{\hat{Q}\hat{Q}}(\omega) 
&= \int_{-\infty}^{\infty} 
\langle\hat{Q}(t)\hat{Q}(0)\rangle e^{i\omega t} \, dt = \frac{1}{2\pi} \int_{-\infty}^{\infty} \langle \hat{Q}(\omega) \hat{Q}(\omega') \rangle \, d\omega' \\[0pt]
&= \chi_{\rm cf}(\omega)\, \chi_{\rm cf}(-\omega)\,
[\mathbf{T}(\omega)]^{T} \mathbf{M}_{\xi} \mathbf{T}(-\omega).
\end{aligned}
\end{equation}

\begin{equation}
\left\langle [\hat{\xi}(\omega)]^{T} 
\hat{\xi}(\omega') \right\rangle
=
\underbrace{\begin{bmatrix}
\bar{n} + \tfrac{1}{2} & 0 & 0 & 0 & 0 & 0 & 0 & 0 & 0 \\[0pt]
0 & \tfrac{1}{2} & \tfrac{i}{2} & 0 & 0 & 0 & 0 & 0 & 0\\[0pt]
0 & -\tfrac{i}{2} & \tfrac{1}{2} & 0 & 0 & 0 & 0 & 0 & 0\\[0pt]
0 & 0 & 0 & \tfrac{1}{2} & \tfrac{i}{2} & 0 & 0 & 0 & 0\\[0pt]
0 & 0 & 0 & -\tfrac{i}{2} & \tfrac{1}{2} & 0 & 0 & 0 & 0\\[0pt]
0 & 0 & 0 & 0 & 0 & \tfrac{1}{2} & \tfrac{i}{2} & 0 & 0\\[0pt]
0 & 0 & 0 & 0 & 0 & -\tfrac{i}{2} & \tfrac{1}{2} & 0 & 0\\[0pt]
0 & 0 & 0 & 0 & 0 & 0 & 0 & S_{\delta\Delta_{h}\delta\Delta_{h}} & 0\\[0pt]
0 & 0 & 0 & 0 & 0 & 0 & 0 & 0 & S_{\delta\Delta_{v}\delta\Delta_{v}}
\end{bmatrix}}_{=:\mathbf{M}_{\xi}}
\, 2\pi \delta(\omega + \omega'),
\end{equation}

\noindent where all optical inputs are coherent or vacuum fields, and $\bar{n}$ is the average thermal phonon occupation.
 
For the detected spectrum, the phase quadrature of the output field from the optical cavity is given by

\begin{align}
\hat{Y}_{h}^{\rm out} = \cos\varphi(-\hat{Y}_{h}^{\rm in} + \sqrt{\kappa^{\rm in}}\hat{Y}_{h}) -\sin\varphi(-\hat{X}_{h}^{\rm in} + \sqrt{\kappa^{\rm in}}\hat{X}_{h}).
\end{align}

\noindent The phase $\varphi$ is determined by the complex amplitude of the output field,

\begin{equation}
e^{-i\varphi} = \frac{h_{\mathrm{out}}}{|h_{\mathrm{out}}|},
\quad 
h_{\rm out} = -h_{\rm in}+\sqrt{\kappa^{\rm in}}h = -\frac{\kappa/2-i\Delta_{h}^{\rm eff}}{\sqrt{\kappa^{\rm in}}} h + \sqrt{\kappa^{\rm in}}h = \frac{-\kappa/2+i\Delta_{h}^{\rm eff}+\kappa^{\rm in}}{\sqrt{\kappa^{\rm in}}}h.
\end{equation}

\noindent The PSD of the detected phase quadrature is then given by

\begin{align}
S_{\hat{Y}_{\rm det}\hat{Y}_{\rm det}} = \eta_{\rm det} S_{\hat{Y}_{h}^{\rm out}\hat{Y}_{h}^{\rm out}} + \frac{1}{2}(1-\eta_{\rm det}),
\end{align}

\noindent where $\eta_{\rm det}$ represents the detection efficiency.

\subsection{The displacement operation}
    
\begin{figure}[htbp]
\centering
\includegraphics[width=.6\linewidth]{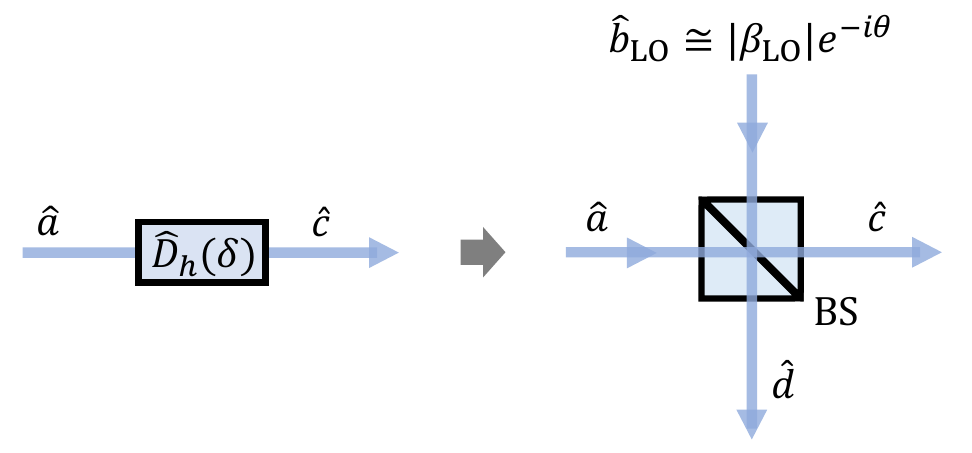}
\caption{The displacement operation $\hat{D}_{h}(\delta)$ with $\hat{a}$ as the input mode and $\hat{c}$ as the output mode. In the experiment, the displacement operation is implemented by combining $\hat{a}$ and a strong local oscillator mode $\hat{b}_{\rm LO}$ at a beamsplitter of transmissivity $T\simeq1$.}
\label{fig:displacement}
\end{figure}

The displacement operation is implemented by combining $\hat{a}$ and a strong local oscillator mode $\hat{b}_{\rm LO}$ at a beamsplitter of transmissivity $T\simeq1$ as shown in Fig. S1. The output mode $\hat{c}$ can be given by:
\begin{equation}
\begin{split}
	\hat{c} &= \sqrt{T}\hat{a} + \sqrt{1-T}e^{i\Psi}\hat{b}_{\mathrm{LO}} \\
 &\simeq  \hat{a} + \underbrace{\sqrt{1-T}|\beta_{\mathrm{LO}}|e^{i(-\theta+\Psi)}}_{=\delta_{0}},
\end{split}
\end{equation}

\noindent where $\Psi \in [-\pi,\pi]$ is a phase offset related to the beamsplitter construction. $\hat{a}$ is the signal beam for the experiment, which is expressed as
\begin{equation}
	\hat{a} = \sqrt{\eta_{\mathrm{i}}}e^{-i\phi}[\hat{h}_{\mathrm{in}}-\sqrt{\kappa^{\mathrm{in}}}\hat{h}]+\sqrt{1-\eta_{\mathrm{i}}}\hat{h}_{\mathrm{i}},
\end{equation}

\noindent where $\hat{h}_{\mathrm{i}}$ is a vacuum noise term considering optical losses between the cavity and the beamsplitter. The angle $x$ is given by

\begin{equation}
	x = \mathrm{Arg}(g_{v}) = \mathrm{Arg}\left[\frac{\sqrt{\kappa^{\text{in}}}}{-i\Delta_{v}^{\text{eff}}+\kappa/2}(\delta + \sqrt{\eta}e^{-i\phi}\left(1-\frac{\kappa^{\text{in}}}{-i\Delta_{h}^{\text{eff}} +\kappa/2}\right)\braket{h_{\text{in}}})\right],
\end{equation}

\noindent where $\delta = \sqrt{\eta_f}\delta_{0}$ and  $\braket{h_{\text{in}}} = \sqrt{P_{\mathrm{in}}/(\hbar\Omega_L)}$, where $P_{\mathrm{in}}$ is the input power of the horizontal mode.

\noindent The angle $x$ is then rewritten as:

\begin{tiny}
\newcommand{\Aa}{\sqrt{\eta_{\mathrm{f}}(1-T)}|\beta_{\mathrm{LO}}|\sin(-\theta)}
\newcommand{\Bb}{\sqrt{\eta_{\mathrm{f}}(1-T)}|\beta_{\mathrm{LO}}|\cos(-\theta)}
\newcommand{\Cc}{\sqrt{\eta}\left(\frac{\kappa^{\rm in}\Delta_{h}^{\rm eff}}{(\kappa/2)^2+(\Delta_{h}^{\rm eff})^{2}}\right)\braket{h_{\text{in}}}}
\newcommand{\Dd}{\sqrt{\eta}\left(1-\frac{\kappa^{\rm in}(\kappa/2)}{(\kappa/2)^2+(\Delta_{h}^{\rm eff})^{2}}\right)\braket{h_{\text{in}}}}
\newcommand{\Ee}{\sqrt{\kappa^{\rm in}}(\kappa/2)}
\newcommand{\Ff}{\sqrt{\kappa^{\rm in}}\Delta_{v}^{\rm eff}}

\begin{align}\label{eq:x-vs-angles}
x = \tan^{-1}\!\left(\frac{\Ee(\Aa+\Cc)+\Ff(\Bb+\Dd)}{\Ee(\Bb+\Dd)-\Ff(\Aa+\Cc)}\right),
\end{align}
\end{tiny}
\noindent where we set $\Psi = 0$ and $\phi = 0$ for simplicity.

\section{Fit based on the model with separable noise sources}

\begin{figure}[htbp]
\centering
\includegraphics[width=.5\linewidth]{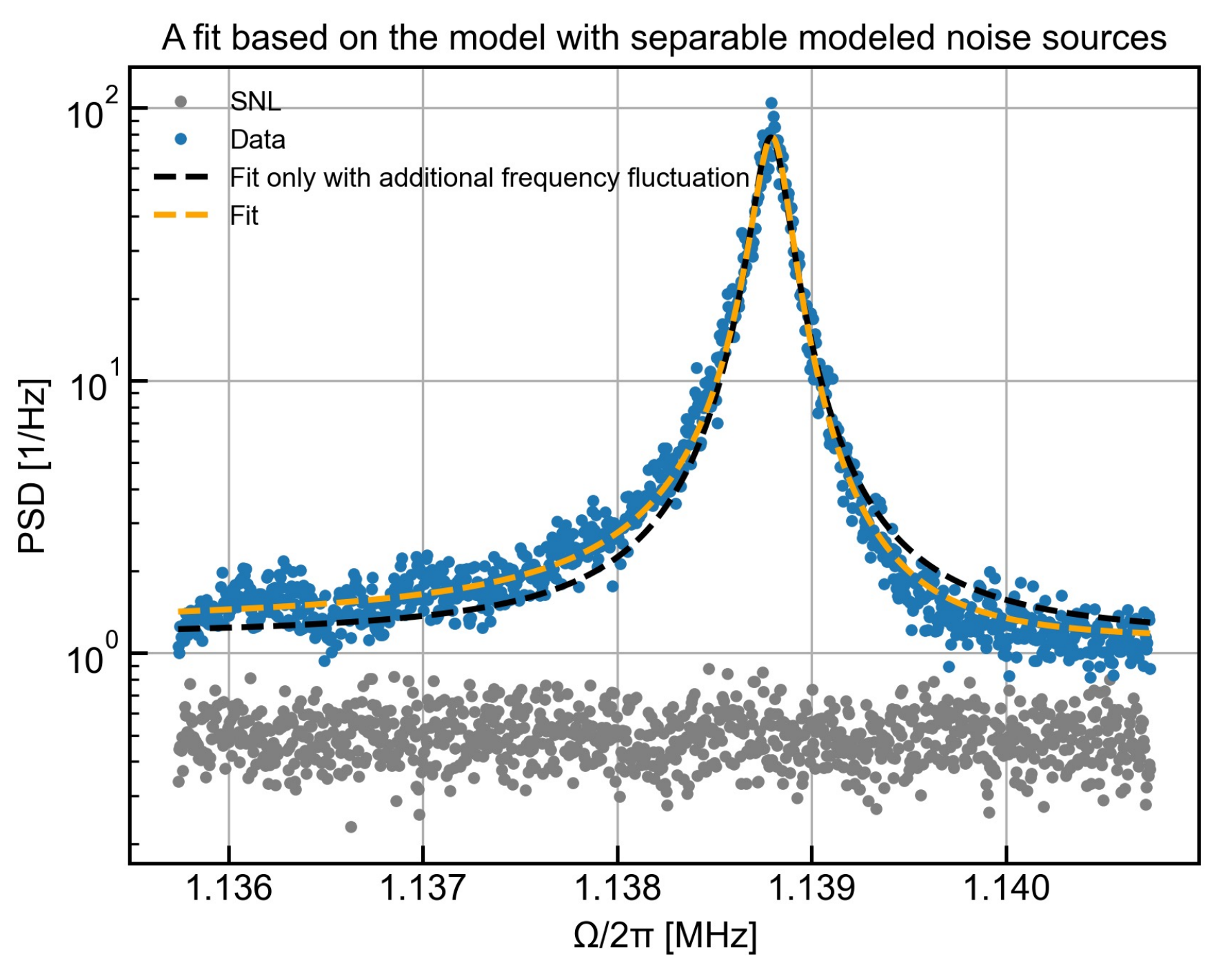}
\caption{Measured spectrum with dynamical backaction cooling and the corresponding fits. The grey dots represent the measured shot-noise limit (SNL) normalized to 0.5. The blue dots show the measured cooling spectrum with $P_{\rm hin} = 50$ $\mu W$ and $P_{\rm vin} = 270$ $\mu W$. The dashed black curve corresponds to a fit including only additional frequency fluctuations ($S_{\delta\Delta\delta\Delta}$), while the dashed orange curve represents the full model incorporating separable noise sources. The fitted parameters are $\Delta_{h}/\kappa = -0.06$, $\Delta_{v}/\kappa = -1.11$, $g_{0,h} = 2\pi\times 3.7$ Hz, $g_{0,v} = 2\pi\times 3.7$ Hz, laser phase noise $S_{\delta\phi\delta\phi} = 1.2\times10^{-12}~[\rm{rad^{2}/Hz}]$, laser amplitude noise $S_{\delta\alpha\delta\alpha} = 40.2$ $\rm{[1/Hz]}$, and additional frequency fluctuation $S_{\delta\Delta_{h}\delta\Delta_{h}} = S_{\delta\Delta_{v}\delta\Delta_{v}} = 0.2~\rm{[Hz^{2}/Hz]}$ at the mechanical frequency.
The full model shows improved agreement with the measured spectrum.}
\label{fig:fit}
\end{figure}

Figure S2 presents an example of fitting the measured spectrum using a dynamical backaction cooling model that separately accounts for classical laser phase noise, amplitude noise, and additional frequency fluctuations, following Sec. 2 of \cite{Daniel24}. This model yields better agreement with the spectral asymmetry and provides phonon numbers consistent with those extracted from our current model. This indicates that representing multiple noise sources by a single effective frequency noise term provides a reasonable simplification. Future work will integrate this modified description into the coherent feedback cooling model for a more comprehensive understanding of the measured spectra.

\section{Experimental Setup}
\begin{figure}
    \centering
    \includegraphics[width=1\linewidth]{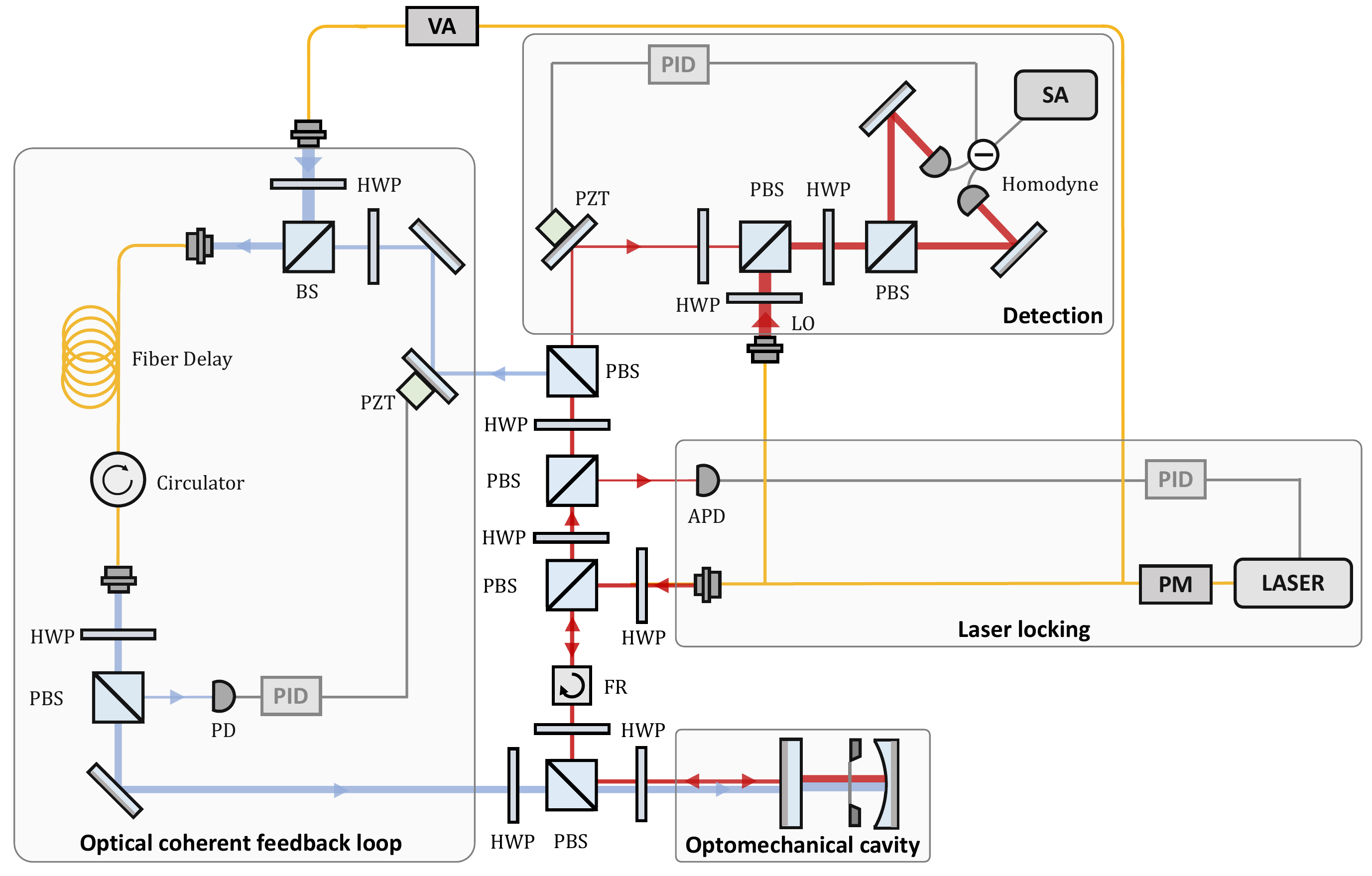}
    \caption{The detailed experimental setup for coherent feedback cooling. PM, phase modulator; PID, proportional–integral–derivative controller; PBS, polarizing beam splitter; HWP, half-wave plate; FR, Faraday rotator; PZT, piezoelectric transducer; VA, variable attenuator; APD, avalanche photodetector; PD, photodiode; BS, beam splitter; LO, local oscillator; SA, spectrum analyzer.}
    \label{fig:SetupS}
\end{figure}

Figure S3 shows the detailed experimental setup of the coherent feedback cooling experiment. The beam from the NKT laser passes through a fiber phase modulator (PM), which modulates the laser at 15 MHz to generate the Pound–Drever–Hall (PDH) sidebands for cavity locking. The beam is then split into three paths: the probe beam, the local oscillator (LO) for homodyne detection, and the auxiliary beam for the displacement operation. The probe beam is rotated to horizontal polarization by a half-wave plate (HWP) before entering the optomechanical cavity, where it interacts with the membrane. The reflected beam is rotated to vertical polarization by a Faraday rotator (FR) and another HWP.  A small portion of the reflected beam ($\approx 2\%$) is sent to an avalanche photodetector (APD), and the PDH error signal is processed by a proportional–integral–derivative (PID) controller that feeds back to the laser frequency actuator to stabilize the laser frequency. Another $\approx 2\%$ of the reflected beam is directed to a homodyne detector for optical readout, where a piezoelectric transducer (PZT) is used for locking the detection angle. The main reflected field is combined with a strong auxiliary beam on a $90:10$ beam splitter to perform the displacement operation required for coherent feedback. The resulting signal is coupled into a fiber delay line and passes through a circulator which filters out the cooling beam reflected from the cavity. After that, a small portion of the delayed signal is tapped out to a photodiode (PD) for stabilizing the relative phase between the signal and the auxiliary beams in the displacement operation. The remaining cooling beam is then re-injected into the cavity. The system parameters are summarized in Table S1.

\begin{table}[h!]
\centering
\caption{\textbf{Experimental parameters.}}
\begin{tabular}{lcc}
\hline\hline
\textbf{Parameter} & \textbf{Symbol} & \textbf{Value} \\
\hline
Mechanical frequency & $\Omega_{\rm m}/2\pi$ & $1.14~\mathrm{MHz}$ \\
Mechanical quality factor & $Q$ & $1.1\times10^8$ \\
Mechanical damping rate & $\Gamma_{\rm m}/2\pi$ & $0.01~\mathrm{Hz}$ \\[0pt]
Cavity linewidth & $\kappa/2\pi$ & $3.7~\mathrm{MHz}$ \\
Cavity finesse & $\mathcal{F}$ & $27{,}000$ \\
Optical wavelength & $\lambda$ & $1549.9~\mathrm{nm}$ \\[0pt]
Optomechanical coupling strength (h mode) & $g_{0,h}/2\pi$ & $3.7~\mathrm{Hz}$ \\
Optomechanical coupling strength (v mode) & $g_{0,v}/2\pi$ & $3.7~\mathrm{Hz}$ \\[0pt]
Input probe power & $P_{\mathrm{hin}}$ & $50~\mu\mathrm{W}$ \\
Input cooling power & $P_{\mathrm{vin}}$ & $270~\mu\mathrm{W}$ \\[0pt]
Escape efficiency & $\eta_{\mathrm{esc}}$ & $68\%$ \\
Cavity mode-matching efficiency (h mode) & $\eta_{\mathrm{mm,h}}$ & $96\%$ \\
Cavity mode-matching efficiency (v mode) & $\eta_{\mathrm{mm,v}}$ & $88\%$ \\
Feedback loop efficiency & $\eta$ & $30\%$ \\
Homodyne detection efficiency & $\eta_{\mathrm{hom}}$ & $90\%$ \\
Total detection efficiency & $\eta_{\mathrm{det}}$ & $1.4\%$ \\[0pt]
\hline\hline
\end{tabular}
\label{tab:exp_parameters}
\end{table}

\section{Phonon number versus areas under simulated spectra}

\begin{figure}[h!]
\centering
\includegraphics[width=.7\linewidth]{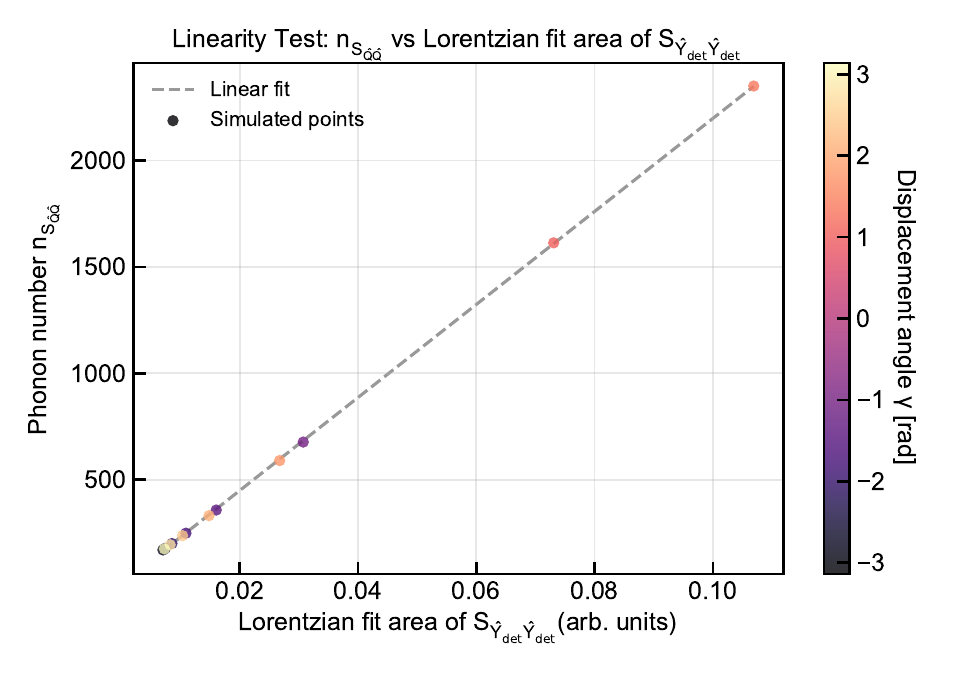}
\caption{Phonon numbers obtained by integrating $S_{\hat{Q}\hat{Q}}$ versus the Lorentzian fit areas of simulated spectra $S_{\hat{Y}_{\rm det}\hat{Y}_{\rm det}}$. $n_{S_{\hat{Q}\hat{Q}}}$ represents the phonon number obtained by integrating $S_{\hat{Q}\hat{Q}}$ with different displacement angles $\gamma$ and with other parameters fixed (see Eq. (7) in the main text). The $X$ axis represents the Lorentzian fit areas below the simulated spectra $S_{\hat{Y}_{\rm det}\hat{Y}_{\rm det}}$. Grey dashed line is a linear fit to simulated points.}
\label{fig:linearity}
\end{figure}

The phonon number of the coherent feedback cooling scheme is obtained by comparing the areas under the PSD with and without coherent feedback cooling, which can be expressed as \cite{Ernzer23} 
\begin{equation}
\bar{n} = \frac{\bar{n}_{\rm calib}}{A_{\rm calib}}A_{\rm det}.
\end{equation}
$\bar{n}_{\rm calib}$ is the phonon number of dynamical backaction cooling scheme, $A_{\rm calib}$ is the corresponding area under the PSD, and $A_{\rm det}$ is the area under the PSD of coherent feedback cooling scheme.
Here we verify the correlation between the phonon number and the PSD area via simulations. Fig. S4 shows the phonon numbers $n_{S_{\hat{Q}\hat{Q}}}$ (obtained by integrating $S_{\hat{Q}\hat{Q}}$) as a function of the Lorentzian fit areas of simulated spectra with different displacement angles $\gamma$ and all other parameters fixed. The grey dashed line indicates a linear correlation, demonstrating that the phonon number scales linearly with the PSD area when the system power and detuning remain constant. This linearity shows that the phonon number in the coherent feedback cooling experiments can be reliably obtained by comparing the areas below the spectra.

\end{document}